\documentclass[11pt,preprint]{elsarticle}
\usepackage[latin9]{inputenc}
\usepackage{array}
\usepackage{multirow}
\usepackage{amsmath}
\usepackage{amssymb}
\usepackage{graphicx}
\usepackage{esint}

\makeatletter

\providecommand{\tabularnewline}{\\}

\usepackage{epsfig}
\usepackage{color}

\makeatother

\begin{document}
\begin{frontmatter} 

\title{Correlation functions for a spin-$\frac{1}{2}$ Ising-XYZ diamond
chain: Further evidence for quasi-phases and pseudo-transitions}

\author[labl1]{I.~M.~Carvalho}

\author[labl2]{J.~Torrico}

\author[labl1]{S.~M.~de~Souza}

\author[labl1]{Onofre~Rojas\corref{cor1}}

\author[labl3,labl1]{Oleg~Derzhko}

\address[labl1]{Departamento de Física, Universidade Federal de Lavras, 37200-000,
Lavras, MG, Brazil}

\address[labl2]{Instituto de Ciências Exatas, Universidade Federal de Alfenas, 37133-840,
Alfenas, MG, Brazil}

\address[labl3]{Institute for Condensed Matter Physics, National Academy of Sciences
of Ukraine, Svientsitskii Street 1, 79011, L'viv, Ukraine}

\cortext[cor1]{email: ors@dfi.ufla.br}

\date{\today}
\begin{abstract}
One-dimensional systems with short-range interactions cannot exhibit
a long-range order at nonzero temperature. However, there are some
particular one-dimensional models, such as the Ising-Heisenberg spin
models with a variety of lattice geometries, which exhibit unexpected
behavior similar to the discontinuous or continuous temperature-driven
phase transition. Although these pseudo-transitions are not true temperature-driven
transitions showing only abrupt changes or sharp peaks in thermodynamic
quantities, they may be confused while interpreting experimental data.
Here we consider the spin-$\frac{1}{2}$ Ising-XYZ diamond chain in
the regime when the model exhibits temperature-driven pseudo-transitions.
We provide a detailed investigation of several correlation functions
between distant spins that illustrates the properties of quasi-phases
separated by pseudo-transitions. Inevitably, all correlation functions
show the evidence of pseudo-transition, which are supported by the
analytical solutions and, besides we provide a rigorous analytical
investigation around the pseudo-critical temperature. It is worth
to mention that the correlation functions between distant spins have
an extremely large correlation length at pseudo-critical temperature. 
\end{abstract}
\end{frontmatter} 

\section{Introduction}

\label{sec1}

In the past decade, Cuesta and Sánchez \cite{Cuesta2004} investigated
relevant properties regarding one-dimensional models with short-range
interaction, such as the general non-existence theorem for finite-temperature
phase transitions \cite{Dyson1969}. Furthermore, there is a wide
class of one-dimensional growth models subjected to an external field
(i.e., with on-site periodic potential), such as the discrete sine-Gordon
model, showing absence of phase transition at finite temperature.
Despite the fact that for the one-dimensional discrete sine-Gordon
model it has been proven that the model cannot have any phase transition
at finite temperature \cite{Cuesta2002}, some numerical simulations
strongly suggested the existence of apparent finite temperature separation
between a flat region and rough phase. This result was investigated
by Ares et al. \cite{Ares2003} using the transfer operator formalism
showing that an arbitrary size sine-Gordon chain will exhibit this
apparent phase transition at finite temperature. Recently, it has
been found that water molecules confined inside single-walled carbon
nanotubes exhibit entirely different behavior from their bulk analogues:
Such single-file chain of water molecules encapsulated in the tubes
shows a temperature-driven quasi-phase transition \cite{Ma2017}.
On the other hand, using the formalism of micro-canonical ensemble
it was also shown a pseudo-transition at finite temperature for a
simple kinetic one-dimensional model \cite{Ferrari1985}.

Lately, several one-dimensional models have been examined in the framework
of decorated structures, in particular, Ising and Heisenberg models
with a variety of geometric structures, such as the Ising-Heisenberg
models in diamond-chain structure \cite{Torrico2014,Torrico2016},
the one-dimensional double-tetrahedral chain, in which the localized
Ising spin regularly alternates with two mobile electrons delocalized
over a triangular plaquette \cite{Galisova2015}, the alternating
Ising-Heisenberg ladder model \cite{Rojas2016}, the Ising-Heisenberg
triangular tube model \cite{Strecka2016}. These models show unexpected
behavior similar to the discontinuous or continuous temperature-driven
phase transition. The analysis of the first derivative of the free
energy, such as entropy, internal energy, magnetization, shows an
abrupt jump as a function of temperature, maintaining a close similarity
with first-order phase transition. Whereas a second order derivative
of free energy, such as specific heat and magnetic susceptibility,
resembles a typical second-order phase transition at finite temperature.
Although these pseudo-transitions are not the true temperature-driven
transitions, abrupt changes or sharp peaks in thermodynamic quantities
may lead to mistaken conclusions while interpreting experimental data.

Here our main goal is to shed further light on pseudo-transitions
and to illustrate them discussing correlation functions around the
pseudo-critical temperature. We take as an example the spin-$\frac{1}{2}$
Ising-XYZ diamond chain investigated in some details earlier \cite{Torrico2014,Torrico2016}.
The rest of the paper is organized as follows. First, we review the
model and its ground-state diagram considered in Refs.~\cite{Torrico2014,Torrico2016,Souza2017},
Sec.~\ref{sec2}. Then we discuss the pseudo-transitions from the
effective Ising-chain-model perspective, Sec.~\ref{sec3}. Our main
findings are the distant pair spin correlation functions for the spin-$\frac{1}{2}$
Ising-XYZ diamond chain, which are examined rigorously in Sec.~\ref{sec4}.
Finally, we summarize our results in Sec.~\ref{sec5}.

\section{Hamiltonian of the model and its ground-state phases}

\label{sec2}

\begin{figure}
\centering{}\includegraphics[scale=0.45]{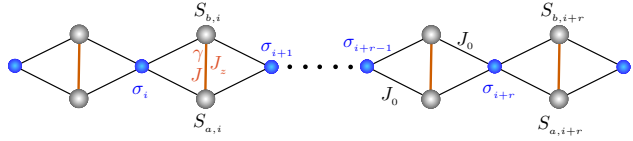} \caption{\label{fig:0} Schematic representation of spin-$\frac{1}{2}$ Ising-XYZ
diamond chain.}
\end{figure}

Here we consider the Hamiltonian of the Ising-XYZ diamond chain, see
Fig.~\ref{fig:0}, as the sum of the block Hamiltonians per unit
cell $\mathcal{H}=\sum_{i=1}^{N}H_{i}$ already discussed in Ref.~\cite{Torrico2014}.
The Hamiltonian of the unit cell is given by: 
\begin{eqnarray}
H_{i}= & - & J(1+\gamma)S_{a,i}^{x}S_{b,i}^{x}-J(1-\gamma)S_{a,i}^{y}S_{b,i}^{y}-J_{z}S_{a,i}^{z}S_{b,i}^{z}\nonumber \\
 & - & J_{0}(S_{a,i}^{z}+S_{b,i}^{z})(\sigma_{i}+\sigma_{i+1})\nonumber \\
 & - & h_{z}(S_{a,i}^{z}+S_{b,i}^{z})-\frac{h}{2}(\sigma_{i}+\sigma_{i+1}),\label{eq:H-orig}
\end{eqnarray}
where $S_{a(b)}^{\alpha}(\alpha=x,y,z)$ are the spin-$\frac{1}{2}$
operators, $\sigma$ corresponds to the Ising spins $\frac{1}{2}$,
$\gamma$ is the $xy$-anisotropy parameter, $J$ and $J_{z}$ are
the Heisenberg-like interactions between interstitial sites, the exchange
parameter $J_{0}$ represents the Ising-like interaction between nodal
and interstitial sites, and the external magnetic field $h_{z}$ and
$h$ are assumed to be along the $z$-direction.

The eigenvalues of the Hamiltonian (\ref{eq:H-orig}) for the $i$-th
unit cell are given by: 
\begin{eqnarray}
\mathcal{E}_{1} & = & -h\frac{\mu}{2}-\frac{J_{z}}{4}+\Delta_{\mu},\label{eq:eps1}\\
\mathcal{E}_{2} & = & -h\frac{\mu}{2}-\frac{J}{2}+\frac{J_{z}}{4},\label{eq:eps2}\\
\mathcal{E}_{3} & = & -h\frac{\mu}{2}+\frac{J}{2}+\frac{J_{z}}{4},\label{eq:eps3}\\
\mathcal{E}_{4} & = & -h\frac{\mu}{2}-\frac{J_{z}}{4}-\Delta_{\mu},\label{eq:eps4}
\end{eqnarray}
where $\mu=\sigma_{i}+\sigma_{i+1}$ and $\Delta_{\mu}=\sqrt{(h_{z}+J_{0}\mu)^{2}+\frac{1}{4}J^{2}\gamma^{2}}$.
The corresponding eigenstates in the natural basis $\{|\begin{smallmatrix}+\\
+
\end{smallmatrix}\rangle,|\begin{smallmatrix}+\\
-
\end{smallmatrix}\rangle,|\begin{smallmatrix}-\\
+
\end{smallmatrix}\rangle,|\begin{smallmatrix}-\\
-
\end{smallmatrix}\rangle\}$ are: 
\begin{eqnarray}
\left|\varphi_{1}\right> & = & -\sin\theta_{\mu}|\begin{smallmatrix}+\\
+
\end{smallmatrix}\rangle+\cos\theta_{\mu}|\begin{smallmatrix}-\\
-
\end{smallmatrix}\rangle,\label{eq:est1}\\
\left|\varphi_{2}\right> & = & \frac{1}{\sqrt{2}}\left(|\begin{smallmatrix}-\\
+
\end{smallmatrix}\rangle+|\begin{smallmatrix}+\\
-
\end{smallmatrix}\rangle\right),\\
\left|\varphi_{3}\right> & = & \frac{1}{\sqrt{2}}\left(|\begin{smallmatrix}-\\
+
\end{smallmatrix}\rangle-|\begin{smallmatrix}+\\
-
\end{smallmatrix}\rangle\right),\\
\left|\varphi_{4}\right> & = & \cos\theta_{\mu}|\begin{smallmatrix}+\\
+
\end{smallmatrix}\rangle+\sin\theta_{\mu}|\begin{smallmatrix}-\\
-
\end{smallmatrix}\rangle,\label{eq:est4}
\end{eqnarray}
where $\theta_{\mu}=\frac{1}{2}\tan^{-1}\frac{J\gamma}{2(h_{z}+J_{0}\mu)}$
with $0<\theta_{\mu}<\pi$.

\begin{figure}
\centering{}\includegraphics[scale=0.6]{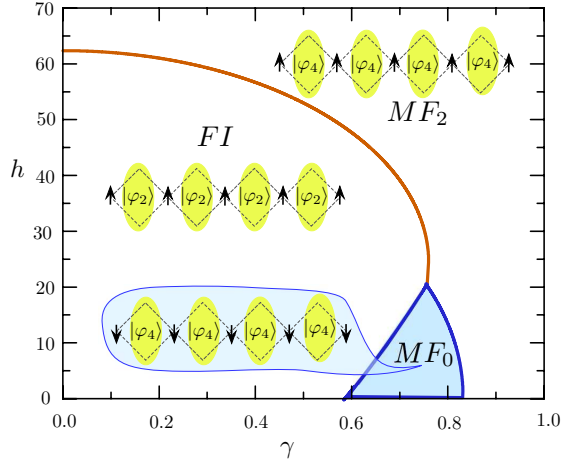} \caption{\label{fig:2} Ground-state phase diagram in the $\gamma-h$ plane.
The coupling parameters are assumed as $J=100$, $J_{z}=24$, and
$J_{0}=-24$.}
\end{figure}

Next, we provide a zero-temperature phase diagram in the $\gamma-h$
plane \cite{Torrico2014,Torrico2016,Souza2017}. Throughout this article,
we will consider only the case $h_{z}=h$. In Fig.~\ref{fig:2} we
report the ground-state phase diagram for a particular set of coupling
parameters $J=100$, $J_{z}=24$, and $J_{0}=-24$. From now on we
will consider just this set of parameters throughout the article.
The phase diagram presents three ground-state phases, namely, one
ferrimagnetic phase ($FI$) and two modulated ferromagnetic Heisenberg
phases ($MF_{0}$ and $MF_{2}$). We use the term ``modulated''
since, e.g., the state $|\varphi_{4}\rangle$ has probability $\cos^{2}\theta_{\mu}$
in $|\begin{smallmatrix}+\\
+
\end{smallmatrix}\rangle$ and $\sin^{2}\theta_{\mu}$ in $|\begin{smallmatrix}-\\
-
\end{smallmatrix}\rangle$. Therefore, these states are given below: 
\begin{alignat}{1}
\left|MF_{2}\right\rangle = & \overset{N}{\underset{i=1}{\prod}}|\varphi_{4}\rangle_{i}\otimes|\uparrow\rangle_{i},\label{eq:state3}\\
\left|FI\right\rangle = & \overset{N}{\underset{i=1}{\prod}}|\varphi_{2}\rangle_{i}\otimes|\uparrow\rangle_{i},\label{eq:state4}\\
\left|MF_{0}\right\rangle = & \overset{N}{\underset{i=1}{\prod}}|\varphi_{4}\rangle_{i}\otimes|\downarrow\rangle_{i}.\label{eq:state2}
\end{alignat}
The corresponding ground-state energies are: 
\begin{alignat}{1}
\varepsilon_{1,0}=E_{_{MF_{2}}}= & -\tfrac{J_{z}}{4}-\tfrac{h}{2}-\!\sqrt{(h_{z}\!+J_{0})^{2}+\tfrac{1}{4}J^{2}\gamma^{2}},\\
\varepsilon_{0,0}=E_{_{FI}}= & \frac{J_{z}}{4}-\frac{J}{2}-\frac{h}{2},\\
\varepsilon_{-1,0}=E_{_{MF_{0}}}= & -\tfrac{J_{z}}{4}+\tfrac{h}{2}-\!\sqrt{(h_{z}\!-J_{0})^{2}+\tfrac{1}{4}J^{2}\gamma^{2}};
\end{alignat}
the $\varepsilon$'s notations should not be confused with $\mathcal{E}$'s
defined in Eqs.~(\ref{eq:eps1}) \textendash{} (\ref{eq:eps4}).
Its worth to note that phases $MF_{2}$ and $MF_{0}$ are degenerate
for a null magnetic field and $\gamma\gtrsim0.5892$. For $\gamma\approx0.5892..0.8314$,
the external magnetic field splits the energy as $E_{_{MF_{2}}}>E_{_{MF_{0}}}$,
whereas for $\gamma\gtrsim0.8314$ the external magnetic field splits
the energy as $E_{_{MF_{2}}}<E_{_{MF_{0}}}$. Further information
about of these results can be found in Refs.~\cite{Souza2017,Torrico2016}.

\section{Effective constants $J_{{\rm {eff}}}$ and $h_{{\rm {eff}}}$ and
pseudo-transitions}

\label{sec3}

Using the decoration transformation \cite{Fisher1959,Syozi1972,Rojas2009,Strecka2010},
we can map the spin-$\frac{1}{2}$ Ising-XYZ diamond chain onto the
well-known spin-$\frac{1}{2}$ Ising chain, whose Hamiltonian is expressed
by $\mathcal{H}_{_{{\rm eff}}}=\sum_{j=1}^{N}\tilde{H}_{j}$, where
\begin{equation}
\mathcal{\tilde{H}}_{j}=-E_{_{{\rm eff}}}^{0}-J_{_{\text{eff}}}\sigma_{j}\sigma_{j+1}-h_{_{\text{eff}}}\sigma_{j},\label{eq:H-eff}
\end{equation}
here $E_{_{{\rm eff}}}^{0}$, $J_{_{\text{eff}}}$, and $h_{_{\text{eff}}}$
are the parameters of the effective Hamiltonian. Through the decoration
transformation \cite{Fisher1959,Syozi1972,Rojas2009,Strecka2010}
these effective Ising-chain parameters can be obtained explicitly;
thus we have 
\begin{alignat}{1}
E_{_{\text{eff}}}^{0}= & \frac{1}{4\beta}\ln\left(w_{1}w_{0}^{2}w_{-1}\right),\\
J_{_{\text{eff}}}= & \frac{1}{\beta}\ln\left(\frac{w_{1}w_{-1}}{w_{0}^{2}}\right),\label{eq:J_eff}\\
h_{_{\text{eff}}}= & \frac{1}{\beta}\ln\left(\frac{w_{1}}{w_{-1}}\right),\label{eq:h_eff}
\end{alignat}
where 
\begin{equation}
w_{\mu}=2\,{\rm e}^{\frac{\beta\mu h}{2}}\left[{\rm e}^{-\frac{\beta J_{z}}{4}}{\rm ch}\tfrac{\beta J}{2}+{\rm e}^{\frac{\beta J_{z}}{4}}{\rm ch}\left(\beta\Delta_{\mu}\right)\right].\label{eq:w_mu}
\end{equation}
Here $\mu=\{-1,0,1\}$, $\beta=\frac{1}{k_{B}T}$, $T$ denotes the
absolute temperature, and $k_{B}$ is the Boltzmann constant.

Along the boundary between $FI$ and $MF_{0}$ or between $FI$ and
$MF_{2}$ the Boltzmann factors \eqref{eq:w_mu} satisfy the following
relation $w_{1}\sim w_{-1}\geqslant w_{0}$, which implies that $w_{1}w_{-1}\geqslant w_{0}^{2}$.
Therefore from Eq.~\eqref{eq:J_eff} we conclude that $J_{{\rm eff}}\geqslant0$
and the equality holds only at $T\rightarrow\infty$. Thus, for $T<\infty$,
the effective parameter is positive $J_{{\rm eff}}>0$ (effective
``ferromagnetic interaction'').

\begin{figure}[h]
\centering{}\includegraphics[scale=0.3]{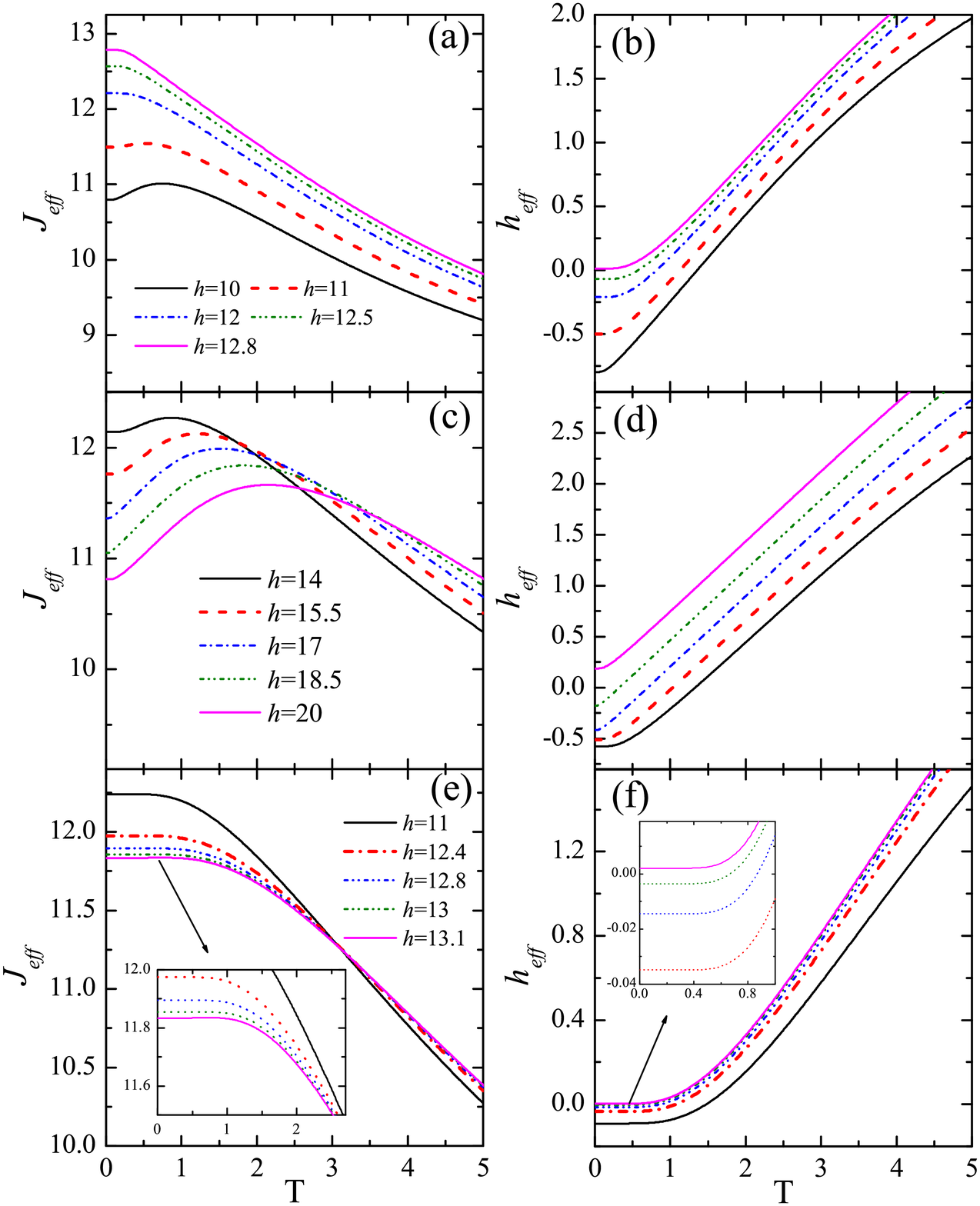} \caption{\label{fig:3} Effective Ising-chain parameters, assuming $J=100$,
$J_{z}=24$, and $J_{0}=-24$. (a) Parameter $J_{{\rm {eff}}}$ as
a function of temperature for $\gamma=0.7$. (b) Effective magnetic
field $h_{{\rm {eff}}}$ as a function of temperature for $\gamma=0.7$.
\label{fig:2b} (c) $J_{{\rm {eff}}}$ against $T$ for $\gamma=0.75$.
(d) $h_{{\rm {eff}}}$ against $T$ for $\gamma=0.75$. \label{fig:2c}
(e) $J_{{\rm {eff}}}$ against $T$ for $\gamma=0.8$. (f) $h_{{\rm {eff}}}$
against $T$ for $\gamma=0.8$.}
\end{figure}

In Fig.~\ref{fig:3}(a), (c), (e), the effective parameter $J_{{\rm eff}}$
\eqref{eq:J_eff} is depicted as a function of temperature for the
above mentioned set of parameters, assuming several values for the
magnetic field $h=h_{z}$. We observe that the effective parameter
$J_{{\rm {eff}}}$ is always ferromagnetic and only weakly depends
on temperature. In Fig.~\ref{fig:3}(b), (d), (f), the effective
magnetic field $h_{_{\text{eff}}}$ \eqref{eq:h_eff} is illustrated.
It is important to remark that the effective magnetic field $h_{{\rm {eff}}}$
may change its sign at a certain temperature, which will be discussed
below.

Let us define the quasi-phases at a finite temperature as the zero-temperature
phase extensions: $MF_{0}$ goes to $qMF_{0}$ and $FI$ goes to $qFI$.
Thereby, in Fig.~\ref{fig:3}(f) one can see an interesting behavior
of $h_{{\rm {eff}}}$ versus temperature $T$, namely, for $\gamma=0.8$
and $h=13$, the effective field $h_{{\rm {eff}}}$ remains almost
zero until $T\approx0.75$. But while $h_{{\rm eff}}<0$ the system
is in $qMF_{0}$ phase, whereas for $h_{{\rm eff}}>0$ the system
goes to $qMF_{2}$ phase, this will be confirmed when we study the
magnetizations of Ising and Heisenberg spins (see Appendix~\ref{sec:app-b}).

Hence, the necessary condition to find pseudo-transition is 
\begin{eqnarray}
h_{{\rm {eff}}}(T_{p})=0, & {\rm and} & w_{0}\ll\{w_{1},w_{-1}\}.\label{eq:h_pc}
\end{eqnarray}
This equation is used to determine the temperature of the pseudo-transition.
In Ref.~\cite{Souza2017}, the equivalent condition, i.e., the requirement
$w_{-1}=w_{1}$, was suggested. It leads to a transcendental equation
for the pseudo-critical temperature $T_{p}$.

This phenomenon contrasts to the ordinary spin-$\frac{1}{2}$ ferromagnetic
Ising chain in a field. The effective magnetic field orders all Ising
spins but as the temperature increases the spins fluctuate and the
ferromagnetic order immediately smoothly melts.

\begin{figure}
\centering{}\includegraphics[scale=0.4]{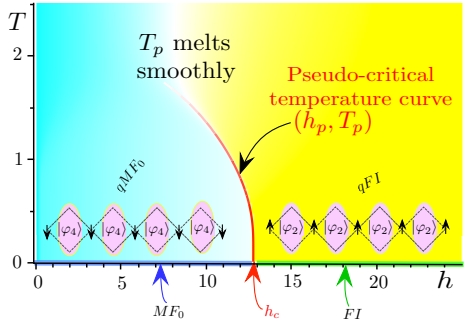} \caption{\label{fig:contour-plot}Phase diagram $T$ against $h$, obtained
from the condition $h_{{\rm eff}}(T_{p})=0$, for fixed parameters
$J=100$, $J_{z}=24$, $J_{0}=-24$, and $\gamma=0.7$.}
\end{figure}

In Fig.~\ref{fig:contour-plot} the pseudo-critical temperature $T_{p}$,
which is determined from the condition $h_{{\rm {eff}}}(T_{p})=0$,
is shown (drawn as a solid red line). $T_{p}$ melts smoothly at $T_{p}\approx1.5$.
The phase diagram clearly shows the pseudo-critical temperature curve
between two regions, the $qMF_{0}$ state and the $qFI$ state for
the model in question with parameters given in Fig.~\ref{fig:contour-plot}.
When $w_{0}$ becomes relevant, the condition $h_{{\rm {eff}}}(T_{p})=0$
still would give, in principle, the value of $T_{p}$, but this result
does not lead to a pseudo-transition because the singularity observed
when $w_{1}=w_{-1}$ vanishes due to the significant contribution
of $w_{0}$. It is also worth mentioning that when $T_{p}\rightarrow0$,
then $h_{p}\rightarrow h_{c}$, where $h_{c}$ is the true critical
magnetic field at the zero temperature.

\begin{table}
\centering{}\caption{\label{tab:P1}Pseudo-critical temperature for a given magnetic field
with the parameters given in Fig.~\ref{fig:contour-plot}. First
two columns correspond to $\gamma=0.7$, second two columns correspond
to $\gamma=0.75$, and third two columns correspond to $\gamma=0.8$.}
\begin{tabular}{|c|c|c|c|c|c|}
\hline 
\multicolumn{2}{|c|}{$\gamma=0.7$} & \multicolumn{2}{c|}{$\gamma=0.75$} & \multicolumn{2}{c|}{$\gamma=0.8$}\tabularnewline
\hline 
\hline 
$h$  & $T_{p}$  & $h$  & $T_{p}$  & $h$  & $T_{p}$\tabularnewline
\hline 
$10$  & $1.3552499$  & $14$  & $1.3552499$  & $11$  & $1.4753981$\tabularnewline
\hline 
$11$  & $1.1270292$  & $15.5$  & $1.1270292$  & $12.4$  & $1.07033229$\tabularnewline
\hline 
$12$  & $0.8150481$  & $17$  & $0.8150481$  & $12.8$  & $0.86683633$\tabularnewline
\hline 
$12.5$  & $0.567641$  & $18.5$  & $0.567641$  & $13$  & $0.66742119$\tabularnewline
\hline 
$12.7$  & $0.3726212$  & $18.7$  & $0.2671694$  & $13.06$  & $0.45524697$\tabularnewline
\hline 
$12.74$  & $0.2694923$  & $18.9$  & $0.2057883$  & $13.0639$  & $0.2973970$\tabularnewline
\hline 
$\mathbf{12.75}$  & $\mathbf{0.0}$  & $\mathbf{19.22}$  & $\mathbf{0.0}$  & $\mathbf{13.063945}$  & $\mathbf{0.0}$\tabularnewline
\hline 
$12.8$  & No $T_{p}$  & $20$  & No $T_{p}$  & $13.1$  & No $T_{p}$\tabularnewline
\hline 
\end{tabular}
\end{table}

In Table~\ref{tab:P1} the pseudo-critical temperature is reported
for several magnetic-field values using the condition \eqref{eq:h_pc}.
Here we assume the fixed $xy$-anisotropy parameters $\gamma=\{0.7,0.75,0.8\}$.
For $\gamma=0.7$ this pseudo-critical temperature occurs in the interface
between $qFI$ and $qMF_{0}$, whereas for $\gamma=0.75$ the pseudo-critical
temperature occurs in the interface of $qFI$, $qMF_{0}$, and $qMF_{2}$.
Analogously, for $\gamma=0.8$ the pseudo-critical temperature occurs
in the boundary between $qMF_{0}$ and $qMF_{2}$. The next-to-last
row of bold data corresponds to the critical field that occurs only
at $T=0$, whereas the last row of data indicates that there is no
pseudo-transition for $h>h_{c}$.

For the considered decorated chain, at some temperature (better low
enough, then there still will be well pronounced traces of the ground-state
ferromagnetic order) the effective magnetic field changes its sign.
All Ising spins reorient simultaneously following the change in the
effective field direction and continue to fluctuate with further temperature
grow.

\section{Spin correlations: Results and discussions}

\label{sec4}

In this section, we study in detail the correlation functions for
the model considered. To this end, we perform an algebraic procedure
discussed in Ref.~\cite{Bellucci2013}. We write the transfer matrix
as follows 
\begin{equation}
\mathbf{W}=\left(\begin{array}{cc}
w_{1} & w_{0}\\
w_{0} & w_{-1}
\end{array}\right),
\end{equation}
where $w_{1}$, $w_{0}$ and $w_{-1}$ are given by \eqref{eq:w_mu}.
Its eigenvalues are expressed by 
\begin{equation}
\lambda_{\pm}=\frac{w_{1}+w_{-1}\pm B}{2}
\end{equation}
with $B=\sqrt{\left(w_{1}-w_{-1}\right)^{2}+4w_{0}^{2}}$. The transfer
matrix $\mathbf{W}$ in the diagonal basis becomes 
\begin{equation}
\left(\begin{array}{cc}
\lambda_{+} & 0\\
0 & \lambda_{-}
\end{array}\right)=\mathbf{P}^{-1}\mathbf{W}\,\mathbf{P}=\boldsymbol{\Lambda},
\end{equation}
where the matrix $\mathbf{P}$ is written as 
\begin{equation}
\mathbf{P}=\left(\begin{array}{cc}
\cos\phi & -\sin\phi\\
\sin\phi & \cos\phi
\end{array}\right)
\end{equation}
with $\phi=\frac{1}{2}\tan^{-1}\frac{2w_{0}}{w_{1}-w_{-1}}$ and $0<\phi<\frac{\pi}{2}$.
Therefore the partition function becomes $Z_{N}=\lambda_{+}^{N}+\lambda_{-}^{N}$
or for large $N$ simply becomes $Z_{N}=\lambda_{+}^{N}$.

Below there are two useful identities that will be used later. Namely,
\begin{alignat}{1}
\cos(2\phi)= & \frac{w_{1}-w_{-1}}{B}=\frac{w_{1}-w_{-1}}{|w_{1}-w_{-1}|}\frac{1}{\sqrt{1+4\bar{w}_{0}^{2}}},\nonumber \\
\sin(2\phi)= & \frac{2w_{0}}{B}=\frac{2\bar{w}_{0}}{\sqrt{1+4\bar{w}_{0}^{2}}}>0,\label{eq:Id-trig}
\end{alignat}
where we define conveniently $\bar{w}_{0}=\frac{w_{0}}{\left|w_{1}-w_{-1}\right|}$.

The expectation value $\langle\sigma\rangle$ is expressed as follows
\begin{equation}
\langle\sigma\rangle=\frac{1}{\lambda_{+}^{N}}{\rm tr}\left(\sigma\mathbf{W}^{N}\right)=\frac{1}{\lambda_{+}^{N}}{\rm tr}\left(\tilde{\sigma}\boldsymbol{\Lambda}^{N}\right),
\end{equation}
where $\tilde{\sigma}=\mathbf{P}^{-1}\sigma\mathbf{P}$ is explicitly
given by 
\begin{equation}
\tilde{\sigma}=\frac{1}{2}\left(\begin{array}{cc}
\cos(2\phi) & -\sin(2\phi)\\
-\sin(2\phi) & -\cos(2\phi)
\end{array}\right).
\end{equation}
After some algebraic manipulations, we obtain 
\begin{equation}
\langle\sigma\rangle=\frac{1}{2}\cos(2\phi)\left(1+u^{N}\right),
\end{equation}
here $u=\tfrac{\lambda_{_{-}}}{\lambda_{_{+}}}$. In the thermodynamic
limit, the formula for $\langle\sigma\rangle$ reduces to 
\begin{alignat}{1}
\langle\sigma\rangle= & \frac{1}{2}\cos(2\phi)=\frac{1}{2}\frac{w_{1}-w_{-1}}{|w_{1}-w_{-1}|}\frac{1}{\sqrt{1+4\bar{w}_{0}^{2}}}.\label{eq:<sg>}
\end{alignat}
The Ising spin magnetization $M_{I}=\langle\sigma\rangle$ close to
the pseudo-transition temperature becomes approximately 
\begin{equation}
M_{I}=\frac{1}{2}\frac{w_{1}-w_{-1}}{|w_{1}-w_{-1}|}\left[1-2\bar{w}_{0}^{2}+\mathcal{O}(\bar{w}_{0}^{4})\right].
\end{equation}
Consequently, the magnetization near the pseudo-transition can be
expressed explicitly as 
\begin{equation}
M_{I}=\begin{cases}
\frac{1}{2}-\bar{w}_{0}^{2}+\mathcal{O}(\bar{w}_{0}^{4}), & w_{1}> w_{-1},\\
-\frac{1}{2}+\bar{w}_{0}^{2}+\mathcal{O}(\bar{w}_{0}^{4}), & w_{1}<w_{-1}.
\end{cases}
\end{equation}

Next we will study the correlation functions. Consider first the thermal
average of two different Ising spins 
\begin{alignat}{1}
\langle\sigma_{j}\sigma_{j+r}\rangle= & \frac{1}{\lambda_{+}^{N}}{\rm tr}\left(\sigma\mathbf{W}^{r}\sigma\mathbf{W}^{N-r}\right)=\frac{1}{\lambda_{+}^{N}}{\rm tr}\left(\tilde{\sigma}\boldsymbol{\Lambda}^{r}\tilde{\sigma}\boldsymbol{\Lambda}^{N-r}\right)
\end{alignat}
with $r=\{0,1,2,\dots\}$. In fact, we are interested in the thermodynamic
limit ($N\rightarrow\infty$) and this equation reduces to 
\begin{alignat}{1}
\langle\sigma_{j}\sigma_{j+r}\rangle & =\frac{1}{4}\left[\cos^{2}(2\phi)+u^{r}\sin^{2}(2\phi)\right]\nonumber \\
 & =\langle\sigma\rangle^{2}+\frac{1}{4}u^{r}\sin^{2}(2\phi)=\langle\sigma\rangle^{2}+\left(\tfrac{w_{0}}{B}\right)^{2}u^{r}.\label{eq:sg1sgr}
\end{alignat}
The case $r=0$ corresponds to the trivial identity $\langle\sigma^{2}\rangle=\frac{1}{4}$.
Now let us discuss the average spin pair \eqref{eq:sg1sgr} around
the pseudo-transition temperature when $\bar{w}_{0}\rightarrow0$.
We have 
\begin{equation}
\langle\sigma_{j}\sigma_{j+r}\rangle=\begin{cases}
\langle\sigma\rangle^{2}+\bar{w}_{0}^{2}\left(\tfrac{w_{-1}}{w_{1}}\right)^{r}+\mathcal{O}\left(\bar{w}_{0}^{4}\right), & w_{1}>w_{-1},\\
\langle\sigma\rangle^{2}+\bar{w}_{0}^{2}\left(\tfrac{w_{1}}{w_{-1}}\right)^{r}+\mathcal{O}\left(\bar{w}_{0}^{4}\right), & w_{1}<w_{-1}.
\end{cases}\label{eq:Ising <sgsg>}
\end{equation}
Therefore, after using \eqref{eq:sg1sgr}, the correlation function
$C_{I}=\langle\sigma_{j}\sigma_{j+r}\rangle-\langle\sigma\rangle^{2}$
becomes 
\begin{equation}
C_{I}=\left(\tfrac{w_{0}}{B}\right)^{2}u^{r},\label{eq:corr-Isng}
\end{equation}
and close to the pseudo-transition ($\bar{w}_{0}\rightarrow0$) the
correlation function reduces to 
\begin{equation}
C_{I}=\begin{cases}
\bar{w}_{0}^{2}\left(\tfrac{w_{-1}}{w_{1}}\right)^{r}+\mathcal{O}\left(\bar{w}_{0}^{4}\right), & w_{1}>w_{-1},\\
\bar{w}_{0}^{2}\left(\tfrac{w_{1}}{w_{-1}}\right)^{r}+\mathcal{O}\left(\bar{w}_{0}^{4}\right), & w_{1}<w_{-1}.
\end{cases}
\end{equation}

\begin{figure}[h]
\centering{}\includegraphics[scale=0.3]{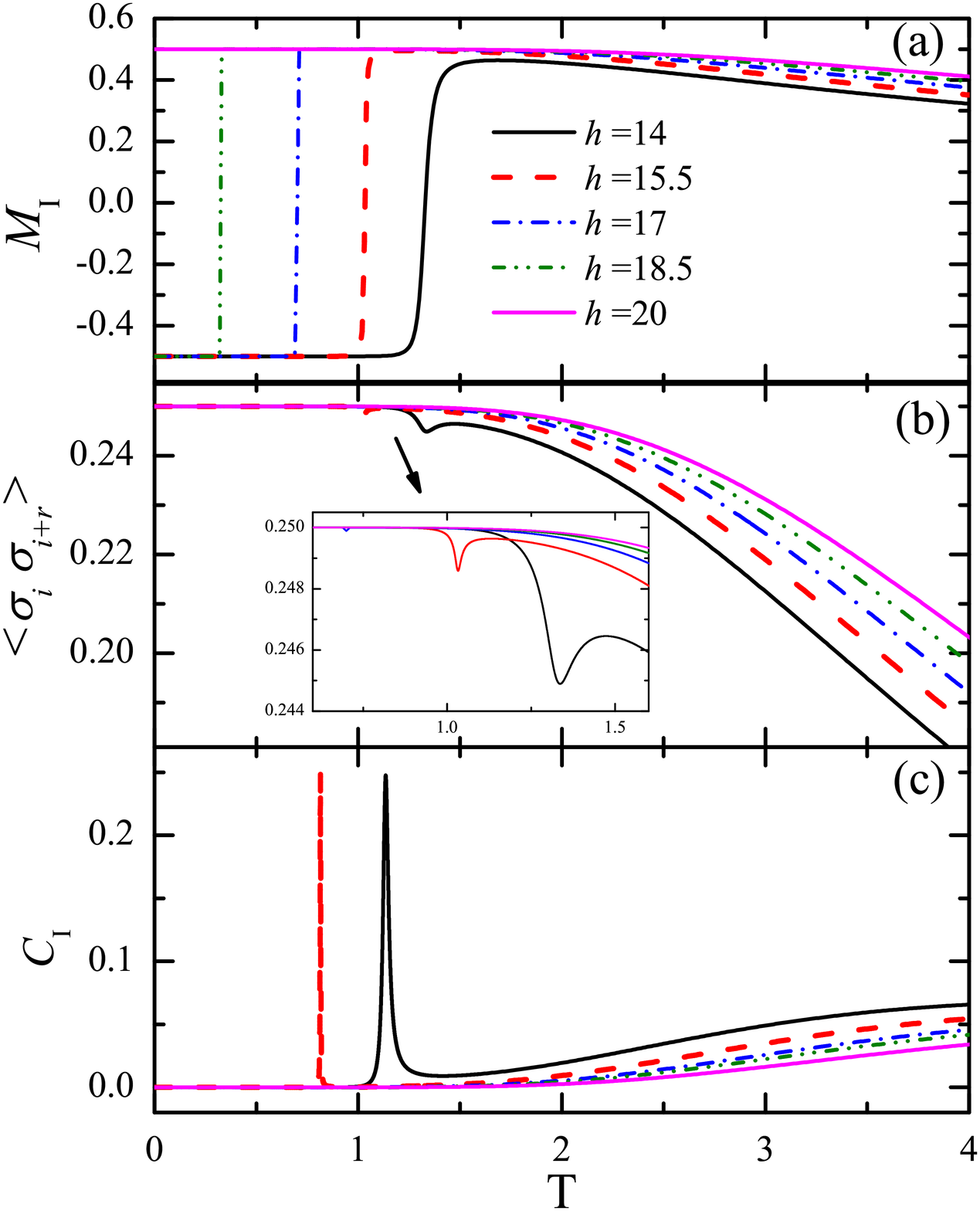} \caption{\label{fig:5}(a) Ising spin magnetization as a function of temperature.
(b) Thermal average $\langle\sigma_{i}\sigma_{i+r}\rangle$, $r=1$
as a function of temperature. (c) Correlation function with $r=1$
as a function of temperature. We assume $J=100$, $J_{z}=24$, $J_{0}=-24$,
and $\gamma=0.75$.}
\end{figure}

In Fig.~\ref{fig:5}(a) we show the temperature dependence of the
Ising spin magnetization. For $w_{1}<w_{-1}$ the magnetization at
$T=0$ is $M_{I}=-\frac{1}{2}$. Similarly, for $w_{1}>w_{-1}$, the
magnetization in the limit of low temperature tends to $M_{I}\rightarrow\frac{1}{2}$,
but holding the condition $w_{1}>w_{-1}$. This change occurs at $h_{{\rm eff}}(T_{p})=0$.
For higher temperatures, the Ising spin magnetization decreases with
temperature, as is expected in ordinary spin models.

In Fig.~\ref{fig:5}(b) the average $\langle\sigma_{j}\sigma_{j+r}\rangle$
is illustrated as a function of temperature. For $T\sim T_{p}$, we
have $\langle\sigma_{j}\sigma_{j+r}\rangle\rightarrow\frac{1}{4}$
for both cases $w_{1}>w_{-1}$ or $w_{1}<w_{-1}$. In principle, it
looks as a monotonically decreasing curve, but as soon as the magnetic
field becomes closer to $h\rightarrow h_{c}$, it exhibits a non-monotonic
temperature behavior suppressed at $T_{p}$ according to Eq.~\eqref{eq:Ising <sgsg>}.

In Fig.~\ref{fig:5}(c) the correlation function \eqref{eq:corr-Isng}
is plotted against the temperature. Here, we can observe a peak at
the pseudo-critical temperature, the lower the temperature is, the
thinner and higher the peak is, whereas the higher the temperature
is, the broader the peak is.

To estimate the magnetization of the Heisenberg spin, we must apply
the decoration transformation approach \cite{Fisher1959,Syozi1972,Rojas2009,Strecka2010}.
First, we have to perform a partial trace over the Heisenberg spins.
For this we need the unit cell Hamiltonian \eqref{eq:H-orig} with
eigenvalues given by Eqs.~\eqref{eq:eps1} \textendash{} \eqref{eq:eps4}
and the corresponding eigenvectors given by Eqs.~\eqref{eq:est1}
\textendash{} \eqref{eq:est4}. With the eigenstates in hand, we can
construct a matrix $\mathbf{Q}$ as follows 
\begin{equation}
\mathbf{Q}=\left(\begin{array}{cccc}
-\sin\theta_{\mu} & 0 & 0 & \cos\theta_{\mu}\\
0 & \frac{1}{\sqrt{2}} & -\frac{1}{\sqrt{2}} & 0\\
0 & \frac{1}{\sqrt{2}} & \frac{1}{\sqrt{2}} & 0\\
\cos\theta_{\mu} & 0 & 0 & \sin\theta_{\mu}
\end{array}\right),
\end{equation}
remembering that $\theta_{\mu}$ was already specified when the eigenstates
(\ref{eq:est1}) \textendash{} (\ref{eq:est4}) were defined. The
matrix $\mathbf{Q}$ diagonalizes the operator $H_{i}$, so any function
of the operator $H_{i}$ will also be diagonalized, hence we have
\begin{equation}
\mathbf{D}=\mathbf{Q}^{-1}{\rm e}^{-\beta H_{i}(\sigma,\sigma')}\mathbf{Q}.
\end{equation}
Consequently, the matrix representation of the operator ${\rm e}^{-\beta H_{i}(\sigma,\sigma')}$
can be written in terms of the eigenstates (\ref{eq:eps1}) \textendash{}
(\ref{eq:eps4}), 
\begin{equation}
\mathbf{D}=\left(\begin{array}{cccc}
{\rm e}^{-\beta\mathcal{E}_{1}} & 0 & 0 & 0\\
0 & {\rm e}^{-\beta\mathcal{E}_{2}} & 0 & 0\\
0 & 0 & {\rm e}^{-\beta\mathcal{E}_{3}} & 0\\
0 & 0 & 0 & {\rm e}^{-\beta\mathcal{E}_{4}}
\end{array}\right).
\end{equation}

The Heisenberg spin operators $S_{a}^{\alpha}$ and $S_{b}^{\alpha}$
can be expressed as $\mathbf{s}_{a}^{\alpha}=S_{a}^{\alpha}\otimes\boldsymbol{1}_{b}$
and $\mathbf{s}_{b}^{\alpha}=\boldsymbol{1}_{a}\otimes S_{b}^{\alpha}$,
respectively, and using the similarity transformation we have $\hat{\mathbf{s}}_{a}^{\alpha}=\mathbf{Q}^{-1}\mathbf{s}_{a}^{\alpha}\mathbf{Q}$
and $\hat{\mathbf{s}}_{b}^{\alpha}=\mathbf{Q}^{-1}\mathbf{s}_{b}^{\alpha}\mathbf{Q}$.
The explicit representation of these matrices is given in Appendix~\ref{sec:app-a}.

In what follows, we perform the partial trace over the Heisenberg
spin operators, 
\begin{alignat}{1}
w_{\mu}^{z}= & {\rm tr}\left(\hat{\mathbf{s}}_{a}^{z}\mathbf{D}\right)={\rm tr}\left(\hat{\mathbf{s}}_{b}^{z}\mathbf{D}\right)\nonumber \\
= & {\rm e}^{\beta\left(\frac{h\mu}{2}+\frac{J_{z}}{4}\right)}\cos(2\theta_{\mu})\sinh(\beta\Delta_{\mu})=\frac{h_{z}+\mu J_{0}}{\Delta_{\mu}}{\rm e}^{\beta\left(\frac{h\mu}{2}+\frac{J_{z}}{4}\right)}\sinh(\beta\Delta_{\mu}).\label{eq:wzmu}
\end{alignat}
The reason why we use the notation $w_{\mu}^{z}$ is because there
is a relation with $w_{\mu}$: 
\begin{equation}
w_{\mu}^{z}=\frac{1}{2\beta}\frac{\partial w_{\mu}}{\partial h_{z}}.\label{eq:dwmu}
\end{equation}
Evidently, using \eqref{eq:dwmu} we can recover the previous result
\eqref{eq:wzmu}. Now we define the matrix $\mathbf{W}_{z}$ as follows
\begin{equation}
\mathbf{W}_{z}=\left(\begin{array}{cc}
w_{1}^{z} & w_{0}^{z}\\
w_{0}^{z} & w_{-1}^{z}
\end{array}\right),
\end{equation}
its elements are given by \eqref{eq:wzmu}.

We are ready now to deal with the expectation values $\langle S_{a}^{z}\rangle$
and $\langle S_{b}^{z}\rangle$ (of course, $\langle S_{a}^{z}\rangle$
and $\langle S_{b}^{z}\rangle$ are identical). Thus, we have 
\begin{equation}
\langle S_{a}^{z}\rangle=\langle S_{b}^{z}\rangle=\langle S^{z}\rangle=\frac{1}{\lambda_{+}^{N}}{\rm tr}\left(\mathbf{W}_{z}\;\mathbf{W}^{N-1}\right).
\end{equation}
Using the similarity transformation given by $\mathbf{\widetilde{W}}_{z}=\mathbf{P}^{-1}\mathbf{W}_{z}\mathbf{P}$,
we can write 
\begin{equation}
\mathbf{\widetilde{W}}_{z}=\left(\begin{array}{cc}
\tilde{w}_{1}^{z} & \tilde{w}_{0}^{z}\\
\tilde{w}_{0}^{z} & \tilde{w}_{-1}^{z}
\end{array}\right).
\end{equation}
Its elements are specified by 
\begin{alignat}{1}
\tilde{w}_{1}^{z}= & w_{1}^{z}\cos^{2}\phi+w_{-1}^{z}\sin^{2}\phi+w_{0}^{z}\sin(2\phi),\\
\tilde{w}_{0}^{z}= & w_{0}^{z}\cos(2\phi)-\frac{1}{2}(w_{1}^{z}-w_{-1}^{z})\sin(2\phi),\\
\tilde{w}_{-1}^{z}= & w_{1}^{z}\sin^{2}\phi+w_{-1}^{z}\cos^{2}\phi-w_{0}^{z}\sin(2\phi).
\end{alignat}
Eliminating the trigonometric functions using Eqs.~\eqref{A8} and
\eqref{A9}, we can also write 
\begin{alignat}{1}
\tilde{w}_{1}^{z}= & \tfrac{\left(w_{1}^{z}+w_{-1}^{z}\right)}{2}+\tfrac{\left(w_{1}^{z}-w_{-1}^{z}\right)\left(w_{1}-w_{-1}\right)}{2B}+\tfrac{2w_{0}^{z}w_{0}}{B},\\
\tilde{w}_{0}^{z}= & \tfrac{w_{0}^{z}\left(w_{1}-w_{-1}\right)}{B}-\tfrac{w_{0}\left(w_{1}^{z}-w_{-1}^{z}\right)}{B},\\
\tilde{w}_{-1}^{z}= & \tfrac{\left(w_{1}^{z}+w_{-1}^{z}\right)}{2}-\tfrac{\left(w_{1}^{z}-w_{-1}^{z}\right)\left(w_{1}-w_{-1}\right)}{2B}-\tfrac{2w_{0}^{z}w_{0}}{B}.
\end{alignat}
Finally, the thermal average of the Heisenberg spin becomes, 
\begin{alignat}{1}
\langle S^{z}\rangle= & \frac{1}{\lambda_{+}^{N}}{\rm tr}\left(\mathbf{\widetilde{W}}_{z}\;\boldsymbol{\Lambda}^{N-1}\right)\nonumber \\
= & \frac{1}{\lambda_{+}^{N}}\left(\tilde{w}_{1}^{z}\lambda_{+}^{N-1}+\tilde{w}_{-1}^{z}\lambda_{-}^{N-1}\right).\label{eq:avr-Sz}
\end{alignat}
We are interested in the thermodynamic limit, so the relation \eqref{eq:avr-Sz}
results in 
\begin{alignat}{1}
\langle S^{z}\rangle= & \frac{\tilde{w}_{1}^{z}}{\lambda_{+}}\nonumber \\
= & \tfrac{\left(w_{1}^{z}+w_{-1}^{z}\right)}{2\lambda_{+}}+\tfrac{\left(w_{1}^{z}-w_{-1}^{z}\right)\left(w_{1}-w_{-1}\right)}{2B\lambda_{+}}+\tfrac{2w_{0}^{z}w_{0}}{B\lambda_{+}}.\label{eq:<sz>}
\end{alignat}
It is worth expressing the average $\langle S^{z}\rangle$ around
the pseudo-critical temperature since this analysis is our main goal.
After performing some algebraic computations, we obtain: 
\begin{alignat}{1}
\langle S^{z}\rangle=\frac{\tilde{w}_{1}^{z}}{\lambda_{+}}= & \frac{h_{z}+J_{0}}{\Delta_{1}w_{1}}{\rm e}^{\beta\left(\frac{h}{2}+\frac{J_{z}}{4}\right)}\sinh(\beta\Delta_{1})\nonumber \\
 & +\frac{2h_{z}}{\Delta_{0}w_{1}}{\rm e}^{\frac{\beta J_{z}}{4}}\sinh(\beta\Delta_{0})\,\bar{w}_{0}+\mathcal{O}\left(\bar{w}_{0}^{2}\right)\label{eq:<sz> wp1>wp1}
\end{alignat}
for $w_{1}>w_{-1}$ and 
\begin{alignat}{1}
\langle S^{z}\rangle=\frac{\tilde{w}_{1}^{z}}{\lambda_{+}}= & \frac{h_{z}-J_{0}}{\Delta_{-1}w_{-1}}{\rm e}^{\beta\left(-\frac{h}{2}+\frac{J_{z}}{4}\right)}\sinh(\beta\Delta_{-1})\nonumber \\
 & +\frac{2h_{z}}{\Delta_{0}w_{-1}}{\rm e}^{\frac{\beta J_{z}}{4}}\sinh(\beta\Delta_{0})\,\bar{w}_{0}+\mathcal{O}\left(\bar{w}_{0}^{2}\right)\label{eq:<sz> wp1>wm1}
\end{alignat}
for $w_{-1}>w_{1}$. The Heisenberg spin magnetization is obtained
from $M_{H}=\langle S^{z}\rangle$. In an analogous way, we can verify
that the magnetizations of Heisenberg spin along the $x$ and $y$-axes
are zero, 
\begin{equation}
\langle S_{a}^{x}\rangle=\langle S_{a}^{y}\rangle=\langle S_{b}^{x}\rangle=\langle S_{b}^{y}\rangle=0.
\end{equation}

Before we pass to computation of the correlation functions between
neighboring cells, we study at first the two distant Heisenberg spins
average, which can be obtained from 
\begin{equation}
\langle S_{j}^{z}S_{j+r}^{z}\rangle=\frac{1}{\lambda_{+}^{N}}{\rm tr}\left(\mathbf{\widetilde{W}}_{z}\;\boldsymbol{\Lambda}^{r-1}\mathbf{\widetilde{W}}_{z}\;\boldsymbol{\Lambda}^{N-r-1}\right),
\end{equation}
here $r=\{1,2,3,\dots\}.$ After some algebraic manipulations and
bearing in mind the thermodynamic limit, we obtain 
\begin{alignat}{1}
\langle S_{j}^{z}S_{j+r}^{z}\rangle= & \frac{\left(\tilde{w}_{1}^{z}\right)^{2}}{\lambda_{+}^{2}}+\frac{\left(\tilde{w}_{0}^{z}\right)^{2}}{\lambda_{+}\lambda_{-}}u^{r}=\langle S^{z}\rangle^{2}+\frac{\left(\tilde{w}_{0}^{z}\right)^{2}}{\lambda_{+}\lambda_{-}}u^{r}.\label{eq:avr-SzSz}
\end{alignat}
Close to the pseudo-critical temperature ($\bar{w}_{0}\rightarrow0$),
the average $\langle S_{j}^{z}S_{j+r}^{z}\rangle$ up to first order
in $\bar{w}$ becomes: 
\begin{alignat}{1}
\langle S_{j}^{z}S_{j+r}^{z}\rangle= & \langle S^{z}\rangle^{2}\!+\!\frac{\left(w_{0}^{z}\right)^{2}}{w_{1}w_{-1}}\negmedspace\left(\!1\!+\!2\frac{w_{-1}^{z}-w_{1}^{z}}{w_{0}^{z}}\bar{w}_{0}\!\right)\negmedspace\left(\frac{w_{-1}}{w_{1}}\right)^{r}\nonumber \\
= & \left(\frac{w_{1}^{z}}{w_{1}}\right)^{2}\left(1+4\frac{w_{0}^{z}}{w_{1}^{z}}\bar{w}_{0}\right)+\frac{\left(w_{0}^{z}\right)^{2}}{w_{1}w_{-1}}\left(1+2\frac{w_{-1}^{z}-w_{1}^{z}}{w_{0}^{z}}\bar{w}_{0}\right)\left(\frac{w_{-1}}{w_{1}}\right)^{r}\nonumber \\
= & \left(\frac{w_{1}^{z}}{w_{1}}\right)^{2}+\frac{\left(w_{0}^{z}\right)^{2}}{w_{1}w_{-1}}\left(\frac{w_{-1}}{w_{1}}\right)^{r}\nonumber \\
 & +\!\frac{2w_{0}^{z}}{w_{1}}\negmedspace\left[\frac{2w_{1}^{z}}{w_{1}}\!+\!\frac{w_{-1}^{z}-w_{1}^{z}}{w_{-1}}\negmedspace\left(\frac{w_{-1}}{w_{1}}\right)^{r}\right]\bar{w}_{0}
\end{alignat}
for $w_{1}>w_{-1}$ and 
\begin{alignat}{1}
\langle S_{j}^{z}S_{j+r}^{z}\rangle= & \left(\frac{w_{-1}^{z}}{w_{-1}}\right)^{2}+\frac{\left(w_{0}^{z}\right)^{2}}{w_{1}w_{-1}}\left(\frac{w_{1}}{w-_{1}}\right)^{r}\nonumber \\
 & \hspace{-0.2cm}+\!\frac{2w_{0}^{z}}{w_{-1}}\negmedspace\left[\!\frac{2w_{-1}^{z}}{w_{-1}}\!+\!\frac{w_{1}^{z}\!-\!w_{-1}^{z}}{w_{1}}\negmedspace\left(\negmedspace\frac{w_{1}}{w-_{1}}\negmedspace\right)^{\negmedspace r}\right]\!\bar{w}_{0}
\end{alignat}
for $w_{-1}>w_{1}$. From \eqref{eq:avr-SzSz} the correlation function
$C_{H}=\langle S_{j}^{z}S_{j+r}^{z}\rangle-\langle S^{z}\rangle^{2}$
becomes 
\begin{equation}
C_{H}=\frac{\left(\tilde{w}_{0}^{z}\right)^{2}}{\lambda_{+}\lambda_{-}}u^{r}.
\end{equation}
Near the pseudo-critical temperature, the correlation function is
expressed by 
\begin{equation}
C_{H}=\begin{cases}
\frac{\left(w_{0}^{z}\right)^{2}}{w_{1}w_{-1}}\negmedspace\left(\!1\!+\!2\frac{w_{-1}^{z}-w_{1}^{z}}{w_{0}^{z}}\bar{w}_{0}\!\right)\negmedspace\left(\frac{w_{-1}}{w_{1}}\right)^{r}, & w_{1}>w_{-1},\\
\frac{\left(w_{0}^{z}\right)^{2}}{w_{1}w_{-1}}\negmedspace\left(\!1\!-\!2\frac{w_{-1}^{z}-w_{1}^{z}}{w_{0}^{z}}\bar{w}_{0}\!\right)\negmedspace\left(\frac{w_{1}}{w-_{1}}\right)^{r}, & w_{1}<w_{-1}.
\end{cases}
\end{equation}

\begin{figure}[h]
\centering{}\includegraphics[scale=0.3]{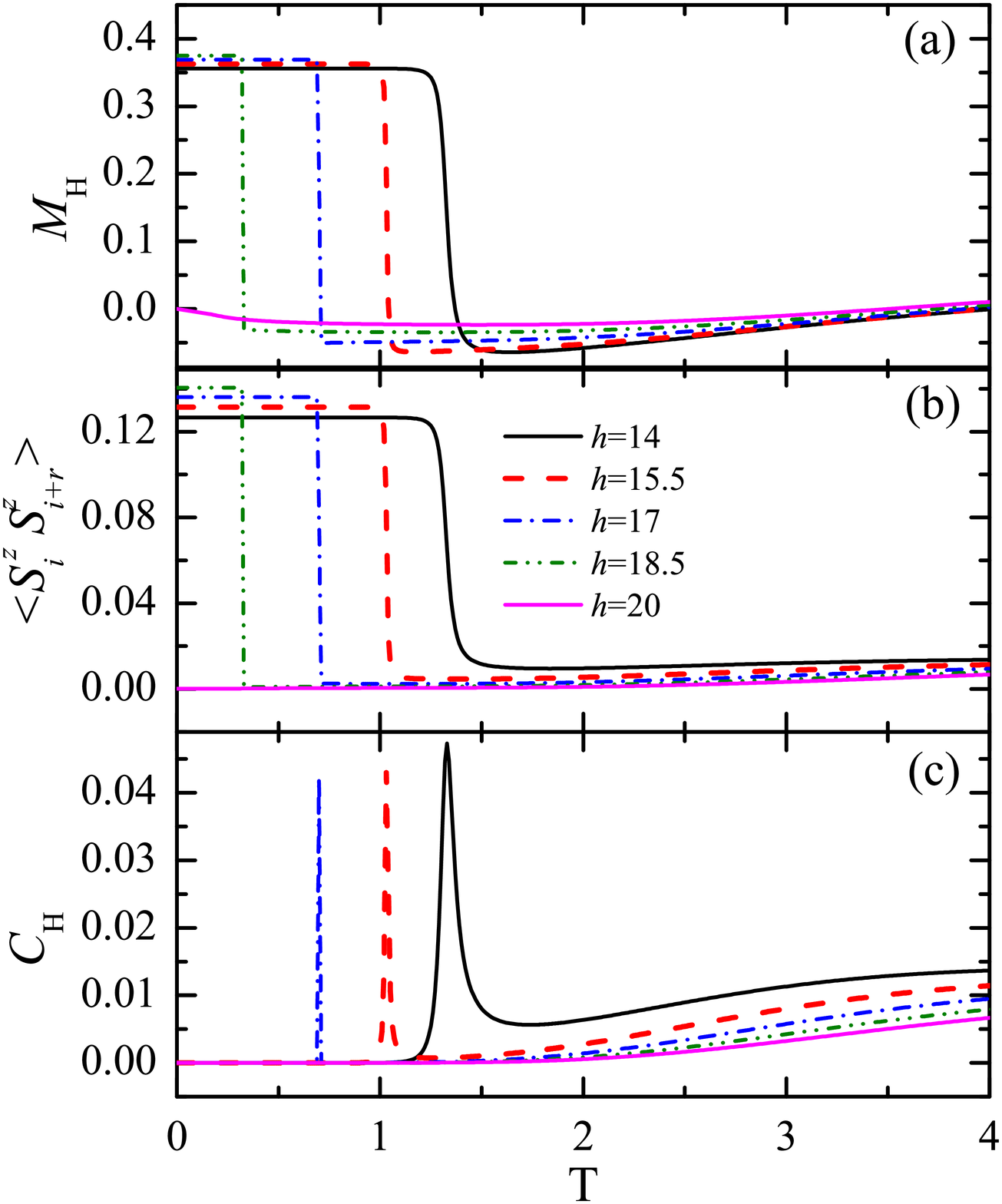} \caption{\label{fig:6}(a) Magnetization per unit cell as a function of temperature.
(b) Average $\langle S_{j}^{z}S_{j+r}^{z}\rangle$, $r=1$ as a function
of temperature. (c) Heisenberg spin correlation function with $r=1$
as a function of temperature. All plots are presented for the fixed
set of parameters $J=100$, $J_{z}=24$, $J_{0}=-24$, and $\gamma=0.75$.}
\end{figure}

In Fig.~\ref{fig:6}(a) we show the temperature dependence of the
Heisenberg spin magnetization for fixed parameters given in the caption
to Fig.~\ref{fig:6} for the range of $h$ values considered in Fig.~\ref{fig:2}.
The behavior of the Heisenberg spin magnetization versus temperature
is in agreement with Eq.~\eqref{eq:<sz>}, and in the limiting cases
close to the pseudo-transition is given by Eqs.~\eqref{eq:<sz> wp1>wp1}
and \eqref{eq:<sz> wp1>wm1}. Obviously, at $T=0$ the magnetization
is in agreement with the ground-state phase diagram {[}the zeroth
order of Eqs.~\eqref{eq:<sz> wp1>wp1} and \eqref{eq:<sz> wp1>wm1}{]}.

In Fig.~\ref{fig:6}(b) the pair Heisenberg spin average $\langle S_{i}^{z}S_{i+r}^{z}\rangle$
is displayed for the same set of parameters. Only the $z$-components
of Heisenberg spins correlate. Below $T_{p}$ the Heisenberg spins
are ordered and $\langle S_{i}^{z}S_{i+r}^{z}\rangle\rightarrow M_{H}^{2}$.
However, for the temperature above $T_{p}$ or at relatively higher
temperatures the correlation functions decreases gradually when $T\to\infty$
as expected for any standard spin chain.

Now we address our investigation to present the quantities which refer
to one cell. So we perform the partial trace over the Heisenberg-Heisenberg
spins operators, 
\begin{alignat}{1}
v_{\mu}^{x}= & {\rm tr}\left(\hat{\mathbf{s}}_{a}^{x}\hat{\mathbf{s}}_{b}^{x}\mathbf{D}\right)=\tfrac{1}{2}{\rm e}^{\frac{\beta\mu h}{2}}\negthickspace\left[{\rm e}^{-\frac{\beta J_{z}}{4}}{\rm sh}\negthickspace\tfrac{\beta J}{2}\!-{\rm e}^{\frac{\beta J_{z}}{4}}{\rm sh}\negthickspace\left(\beta\Delta_{\mu}\!\right)\!\cos(2\theta_{\mu})\right]\negthickspace,\\
v_{\mu}^{y}= & {\rm tr}\left(\hat{\mathbf{s}}_{a}^{y}\hat{\mathbf{s}}_{b}^{y}\mathbf{D}\right)=\tfrac{1}{2}{\rm e}^{\frac{\beta\mu h}{2}}\negthickspace\left[{\rm e}^{-\frac{\beta J_{z}}{4}}{\rm sh}\negthickspace\tfrac{\beta J}{2}\!+{\rm e}^{\frac{\beta J_{z}}{4}}{\rm sh}\negthickspace\left(\beta\Delta_{\mu}\!\right)\!\cos(2\theta_{\mu})\right]\negthickspace,\\
v_{\mu}^{z}= & {\rm tr}\left(\hat{\mathbf{s}}_{a}^{z}\hat{\mathbf{s}}_{b}^{z}\mathbf{D}\right)=\tfrac{1}{2}{\rm e}^{\frac{\beta\mu h}{2}}\left[{\rm e}^{\frac{\beta J_{z}}{4}}{\rm ch}\left(\beta\Delta_{\mu}\right)-{\rm e}^{-\frac{\beta J_{z}}{4}}{\rm ch}\left(\tfrac{\beta J}{2}\right)\right].
\end{alignat}
Therefore, the corresponding matrix can be written as 
\begin{equation}
\mathbf{V}_{\alpha}=\left(\begin{array}{cc}
v_{1}^{\alpha} & v_{0}^{\alpha}\\
v_{0}^{\alpha} & v_{-1}^{\alpha}
\end{array}\right),
\end{equation}
with $\alpha=\{x,y,z\}$. Using the similarity transformation given
by $\mathbf{\widetilde{V}}_{\alpha}=\mathbf{P}^{-1}\mathbf{V}_{\alpha}\mathbf{P}$,
we have 
\begin{equation}
\mathbf{\tilde{V}}_{\alpha}=\left(\begin{array}{cc}
\tilde{v}_{1}^{\alpha} & \tilde{v}_{0}^{\alpha}\\
\tilde{v}_{0}^{\alpha} & \tilde{v}_{-1}^{\alpha}
\end{array}\right),
\end{equation}
with the matrix elements 
\begin{alignat}{1}
\tilde{v}_{1}^{\alpha}= & v_{1}^{\alpha}\cos^{2}\phi+v_{-1}^{\alpha}\sin^{2}\phi+v_{0}^{\alpha}\sin(2\phi),\\
\tilde{v}_{0}^{\alpha}= & v_{0}^{\alpha}\cos(2\phi)-\frac{1}{2}(v_{1}^{\alpha}-v_{-1}^{\alpha})\sin(2\phi),\\
\tilde{v}_{-1}^{\alpha}= & v_{1}^{\alpha}\sin^{2}\phi+v_{-1}^{\alpha}\cos^{2}\phi-v_{0}^{\alpha}\sin(2\phi).
\end{alignat}
By eliminating the trigonometric functions, the elements of matrix
$\mathbf{\tilde{V}}_{\alpha}$ become 
\begin{alignat}{1}
\tilde{u}_{1}^{\alpha}= & \tfrac{\left(u_{1}^{\alpha}+u_{-1}^{\alpha}\right)}{2}+\tfrac{\left(u_{1}^{\alpha}-u_{-1}^{\alpha}\right)\left(w_{1}-w_{-1}\right)}{2B}+\tfrac{2u_{0}^{\alpha}w_{0}}{B},\\
\tilde{u}_{0}^{\alpha}= & \tfrac{u_{0}^{\alpha}\left(w_{1}-w_{-1}\right)}{B}-\tfrac{w_{0}\left(u_{1}^{\alpha}-u_{-1}^{\alpha}\right)}{B},\\
\tilde{u}_{-1}^{\alpha}= & \tfrac{\left(u_{1}^{\alpha}+u_{-1}^{\alpha}\right)}{2}-\tfrac{\left(u_{1}^{\alpha}-u_{-1}^{\alpha}\right)\left(w_{1}-w_{-1}\right)}{2B}-\tfrac{2u_{0}^{\alpha}w_{0}}{B}.
\end{alignat}
Consequently, the average of the one-cell pair of Heisenberg spins
is given by 
\begin{alignat}{1}
\langle S_{a,j}^{\alpha}S_{b,j}^{\alpha}\rangle= & \frac{1}{\lambda_{+}^{N}}{\rm tr}\left(\mathbf{\widetilde{V}}_{\alpha}\;\boldsymbol{\Lambda}^{N-1}\right)=\frac{1}{\lambda_{+}^{N}}\left(\tilde{v}_{1}^{\alpha}\lambda_{+}^{N-1}+\tilde{v}_{-1}^{\alpha}\lambda_{-}^{N-1}\right).\label{eq:avr-S_alf}
\end{alignat}
In the thermodynamic limit, we have 
\begin{alignat}{1}
\langle S_{a,j}^{\alpha}S_{b,j}^{\alpha}\rangle\!= & \tfrac{\left(v_{1}^{\alpha}+v_{-1}^{\alpha}\right)}{2\lambda_{+}}+\tfrac{\left(v_{1}^{\alpha}-v_{-1}^{\alpha}\right)\left(w_{1}-w_{-1}\right)}{2B\lambda_{+}}+\tfrac{2v_{0}^{\alpha}w_{0}}{B\lambda_{+}}.\label{eq:SS_alph}
\end{alignat}
In addition, the averages $\langle S_{a}^{x}S_{b}^{x}\rangle$, $\langle S_{a}^{y}S_{b}^{y}\rangle$,
and $\langle S_{a}^{z}S_{b}^{z}\rangle$ satisfy the following identity
\begin{eqnarray}
\langle S_{a}^{x}S_{b}^{x}\rangle+\langle S_{a}^{y}S_{b}^{y}\rangle+\langle S_{a}^{z}S_{b}^{z}\rangle=\frac{1}{4}
\end{eqnarray}
at any temperature. This is a simple consequence of the obvious relation
$\langle({\bf {S}}_{a}+{\bf {S}}_{b})^{2}\rangle=S(S+1)=2$.

\begin{figure}
\centering{}\includegraphics[scale=0.3]{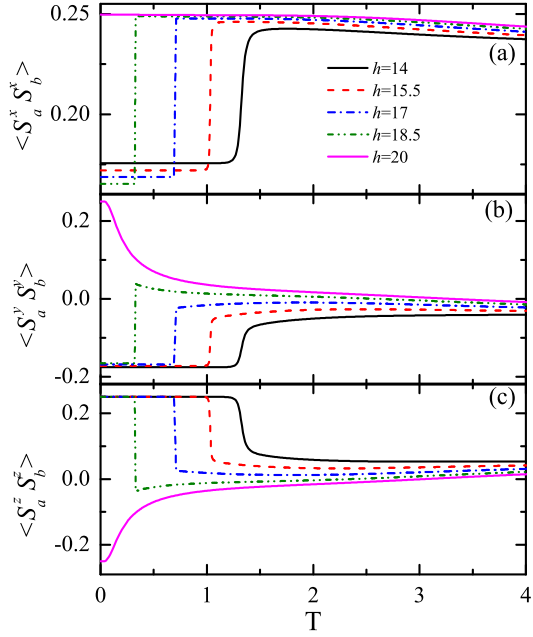} \caption{\label{fig:1-cell-corr}One-cell pair average ($r=0$) as a function
of temperature. (a) $\langle S_{a}^{x}S_{b}^{x}\rangle$; (b) $\langle S_{a}^{y}S_{b}^{y}\rangle$;
(c) $\langle S_{a}^{z}S_{b}^{z}\rangle$. We assume $J=100$, $J_{z}=24$,
$J_{0}=-24$, and $\gamma=0.75$.}
\end{figure}

In Fig.~\ref{fig:1-cell-corr}(a) the average of one-cell of $\langle S_{a}^{x}S_{b}^{x}\rangle$
is reported as a function of temperature for a range of magnetic field
described in panel (a) and considering the same set of parameters
as in Fig.~\ref{fig:6}. For temperatures $0\le T\lesssim T_{p}$
the average $\langle S_{a}^{x}S_{b}^{x}\rangle$ is around a certain
value definitely below $\frac{1}{4}$ and above $T_{p}$ it becomes
almost $\frac{1}{4}$ and then decreases with further temperature
growth. Please, check the statement. Analogously, Fig.~\ref{fig:1-cell-corr}(b)
refers to $\langle S_{a}^{y}S_{b}^{y}\rangle$, where there is also
a clear jump at $T=T_{p}$. The lower the temperature is, the jump
becomes more evident. In Fig.~\ref{fig:1-cell-corr}(c), $\langle S_{a}^{z}S_{b}^{z}\rangle$
is depicted as a function of temperature. For temperatures lower than
$T_{p}$, $\langle S_{a}^{z}S_{b}^{z}\rangle\rightarrow\frac{1}{4}$.
Whereas for $h>h_{c}$ there is no pseudo-transition and $\langle S_{a}^{z}S_{b}^{z}\rangle\rightarrow-\frac{1}{4}$
when $T\rightarrow0$.

Furthermore, let us consider the thermal average between different
mixed Ising spin and Heisenberg spin at distant sites, 
\begin{equation}
\langle S_{j}^{z}\sigma_{j+r}\rangle=\frac{1}{\lambda_{+}^{N}}{\rm tr}\left(\mathbf{\widetilde{W}}_{z}\;\boldsymbol{\Lambda}^{r-1}\tilde{\sigma}\;\boldsymbol{\Lambda}^{N-r}\right),
\end{equation}
here we assume $r=\{1,2,3,\ldots\}$. After algebraic manipulations
similar to the previous case, and taking the thermodynamic limit,
we obtain the following expression, 
\begin{equation}
\langle S_{j}^{z}\sigma_{j+r}\rangle=\frac{\tilde{w}_{1}^{z}}{2\lambda_{+}}\cos(2\phi)-\frac{\tilde{w}_{0}^{z}}{2\lambda_{-}}\sin(2\phi)u^{r}.\label{eq:avr-Szsg}
\end{equation}
Writing it in terms of the Boltzmann factors, we have 
\begin{equation}
\langle S_{j}^{z}\sigma_{j+r}\rangle=\langle S^{z}\rangle\langle\sigma\rangle-\frac{\tilde{w}_{0}^{z}w_{0}}{B\lambda_{-}}u^{r}.\label{eq:avr-Szsg-alt}
\end{equation}
Around the pseudo-critical temperature, the previous result reduces
to 
\begin{equation}
\langle S_{j}^{z}\sigma_{j+r}\rangle\!=\frac{w_{1}^{z}}{2w_{1}}+\left[\frac{w_{0}^{z}}{w_{1}}-\frac{w_{0}^{z}}{w_{-1}}\left(\frac{w_{-1}}{w_{1}}\right)^{r}\right]\bar{w}_{0}+\mathcal{O}(\bar{w}_{0}^{2})\label{eq:Ssg-wp1>wm1}
\end{equation}
for $w_{1}>w_{-1}$ and to 
\begin{equation}
\langle S_{j}^{z}\sigma_{j+r}\rangle\!=\frac{w_{-1}^{z}}{2w_{-1}}+\left[\frac{w_{0}^{z}}{w_{-1}}-\frac{w_{0}^{z}}{w_{1}}\left(\frac{w_{1}}{w_{-1}}\right)^{r}\right]\bar{w}_{0}+\mathcal{O}(\bar{w}_{0}^{2})\label{eq:Ssg-wp1<wm1}
\end{equation}
for $w_{1}<w_{-1}$. From \eqref{eq:avr-Szsg} we obtain the correlation
function 
\begin{equation}
C_{IH}=-\frac{\tilde{w}_{0}^{z}}{2\lambda_{-}}\sin(2\phi)u^{r}=-\frac{\tilde{w}_{0}^{z}w_{0}}{B\,\lambda_{-}}u^{r},
\end{equation}
and near the pseudo-critical temperature ($\bar{w}_{0}\rightarrow0$)
we have 
\begin{equation}
C_{IH}=\begin{cases}
\frac{w_{0}^{z}}{w_{-1}}\left(\frac{w_{-1}}{w_{1}}\right)^{r}\bar{w}_{0}+\mathcal{O}(\bar{w}_{0}^{2}), & w_{1}>w_{-1},\\
\frac{w_{0}^{z}}{w_{1}}\left(\frac{w_{1}}{w_{-1}}\right)^{r}\bar{w}_{0}+\mathcal{O}(\bar{w}_{0}^{2}), & w_{1}<w_{-1}.
\end{cases}\label{eq:C_IH}
\end{equation}

A similar algebraic computation was developed in Ref.~\cite{Bellucci2013},
but here we concentrate on the case near the pseudo-critical temperature.

\begin{figure}
\centering{}\includegraphics[scale=0.3]{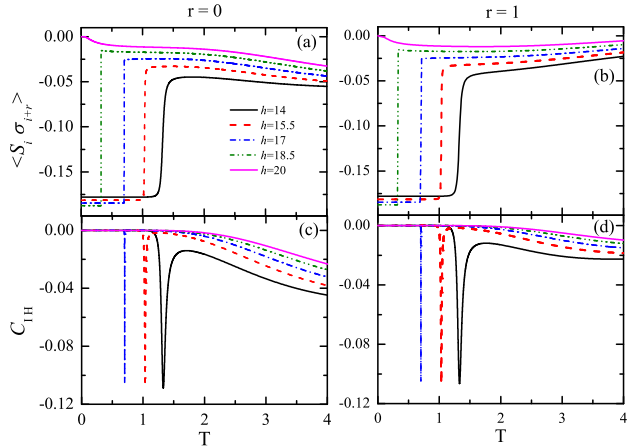} \caption{\label{fig:8}(a) and (b) Average $\langle S_{i}^{z}\sigma_{i+1}\rangle$
as a function of temperature. (c) and (d) Correlation function as
a function of temperature. We assume fixed $J=100$, $J_{z}=24$,
$J_{0}=-24$, and $\gamma=0.75$.}
\end{figure}

In Fig.~\ref{fig:8}, we show the Ising-Heisenberg spin pair average
and the Ising-Heisenberg spin correlation function as a function of
temperature, assuming the same set of parameters as in Fig.~\ref{fig:6}
and the values of $h$ as given in the legend of Fig.~\ref{fig:8}.
Panel (a) illustrates the one-cell pair Ising-Heisenberg spin average
($r=0$, see Fig.~\ref{fig:0}). We observe here a noticeable jump
around the pseudo-critical temperature $T_{p}$. The same quantity
is displayed in panel (b) but now for $r=1$. In panel (c), the correlation
function for one-cell Ising-Heisenberg spin pair is depicted, where
we observe a strong depressing of the curves at pseudo-critical temperature
$T_{p}$. Similarly in panel (d) the correlation function for $r=1$
is illustrated, and analogous behavior is observed. Therefore, all
represented curves are entirely in agreement with the average of one-cell
Ising-Heisenberg spin pair \eqref{eq:Ssg-wp1>wm1} and \eqref{eq:Ssg-wp1<wm1},
as well as with the correlation function provided by Eq.~\eqref{eq:C_IH}.

\begin{figure}[h]
\centering{}\includegraphics[scale=0.3]{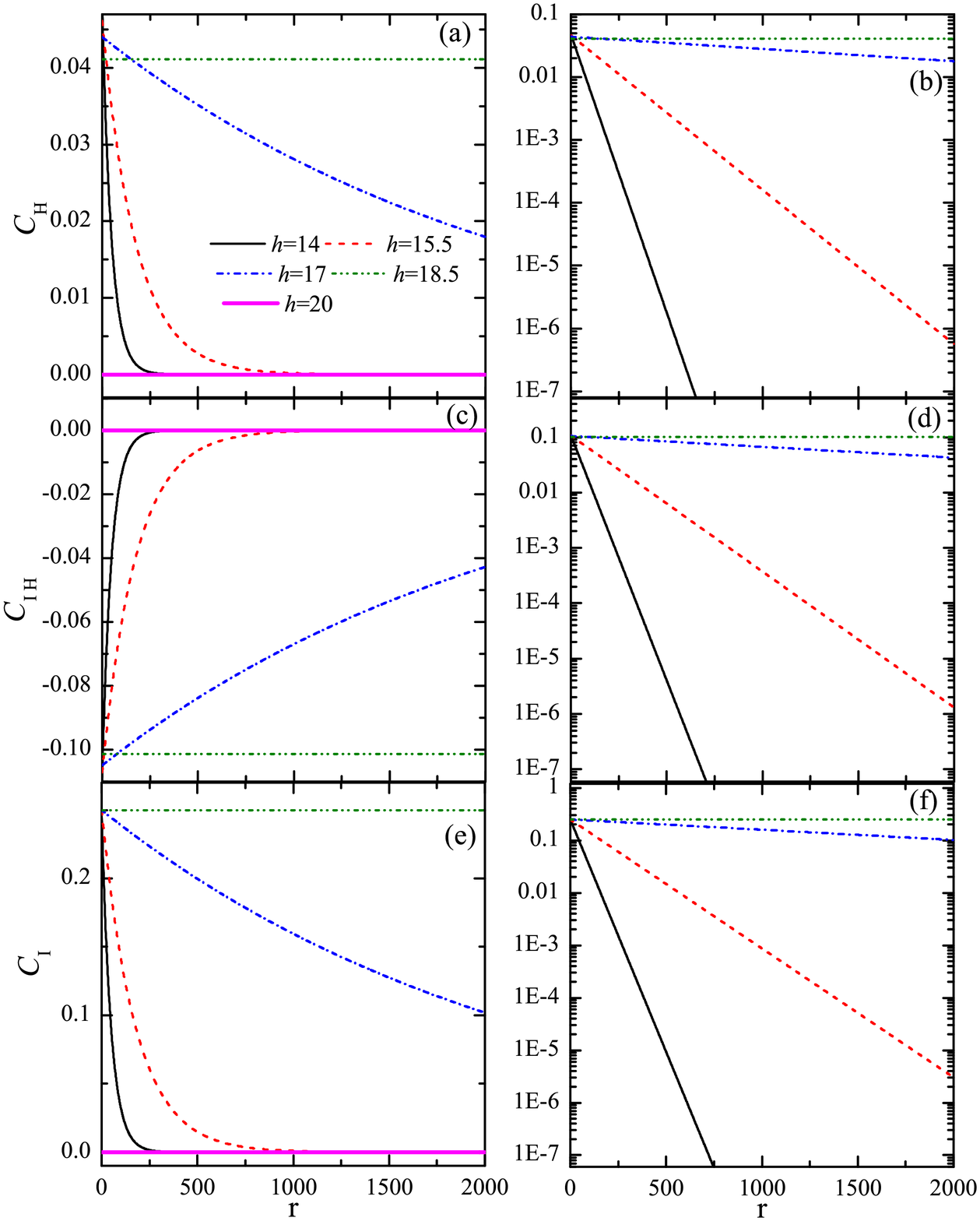} \caption{\label{fig:9}Correlation function decay with distance $r$ for $J=100$,
$J_{z}=24$, $J_{0}=-24$, and $\gamma=0.75$. (a) For $\langle S_{i}^{z}S_{i+r}^{z}\rangle-\langle S_{i}^{z}\rangle\langle S_{i}^{z}\rangle$
against $r$. (b) Same as (a) but in logarithmic scale. (c) $\langle S_{i}^{z}\sigma_{i+r}\rangle-\langle S_{i}^{z}\rangle\langle\sigma_{i}\rangle$
against $r$. (d) Same as (c) but in logarithmic scale. (e) $\langle\sigma_{i}\sigma_{i+1}\rangle-\langle\sigma_{i}\rangle^{2}$
against $r$. (f) Same as (e) but in logarithmic scale.}
\end{figure}

Now we discuss the dependence of correlation functions on the interspin
distance for several values of the magnetic field (and the corresponding
pseudo-critical temperatures, see Fig.~\ref{fig:contour-plot} and
Table~\ref{tab:P1}) which are shown in Fig.~\ref{fig:9}. From
panel (a) it can be seen that the correlation function between Heisenberg-Heisenberg
spins $C_{H}$ for low magnetic fields decays significantly with the
increase of distance $r$, but while $h\rightarrow h_{c}$ ($h<h_{c}$)
the decay becomes less significant, that means the correlations are
strong even for far away spins. For example, for $h=18.5$ the correlation
function becomes almost independent of distance $r$ up to $r\approx2000$
(for more details see Table \ref{tab:P2}). The same plot is depicted
in panel (b), but the correlation function is given in logarithmic
scale here. It is simply a straight line with a slope $\ln\frac{\lambda_{-}}{\lambda_{+}}$,
which is obviously negative because $\lambda_{-}<\lambda_{+}$. However,
for $h=18.5$ this slopes is almost zero, while for lower magnetic
fields the module of these slopes are large. Basically similar plots
are shown in Fig.~\ref{fig:9}(c), (d) for distant mixed Ising and
Heisenberg spins correlation functions $C_{IH}$; these correlation
functions are negative. The distant pair Ising-Ising spin correlations
$C_{I}$, illustrated in Fig.~\ref{fig:9}(e), (f), are positive
and show the same behavior as the Heisenberg-Heisenberg correlation
functions. Certainly, we also observe in all panels the zero correlation
functions for $h>h_{c}$.

\begin{table}
\centering{}\caption{\label{tab:P2}Correlation function decay with distance. Here we report
the ratio of $\frac{C(r)}{C(1)}$ in percent at the pseudo-critical
temperature $T_{p}$. The three types of the correlation functions
$C_{I}$, $C_{H}$ and $C_{IH}$ are denoted by $C$. We assume the
fixed $xy$-anisotropy parameter $\gamma=\{0.7,0.75,0.8\}$.}
\begin{tabular}{|c|c|c|}
\hline 
\multicolumn{3}{|c|}{$\gamma=0.7$}\tabularnewline
\hline 
\hline 
$h$ & $r$ & $\frac{C(r)}{C(1)}[\%]$\tabularnewline
\hline 
\hline 
\multirow{3}{*}{$10$} & $20$ & $50.15$\tabularnewline
\cline{2-3} 
 & $50$ & $11.73$\tabularnewline
\cline{2-3} 
 & $100$ & $2.75$\tabularnewline
\hline 
\multirow{3}{*}{$11$} & $50$ & $53.35$\tabularnewline
\cline{2-3} 
 & $100$ & $28.10$\tabularnewline
\cline{2-3} 
 & $300$ & $2.16$\tabularnewline
\hline 
\multirow{3}{*}{$12$} & $500$ & $53.05$\tabularnewline
\cline{2-3} 
 & $1000$ & $28.11$\tabularnewline
\cline{2-3} 
 & $3000$ & $2.21$\tabularnewline
\hline 
\multirow{3}{*}{$12.5$} & $1\negthickspace\times\negthickspace10^{4}$ & $69.78$\tabularnewline
\cline{2-3} 
 & $5\negthickspace\times\negthickspace10^{4}$ & $16.55$\tabularnewline
\cline{2-3} 
 & $1\negthickspace\times\negthickspace10^{5}$ & $2.74$\tabularnewline
\hline 
\end{tabular}%
\begin{tabular}{|c|c|c|}
\hline 
\multicolumn{3}{|c|}{$\gamma=0.75$}\tabularnewline
\hline 
\hline 
$h$ & $r$ & $\frac{C(r)}{C(1)}[\%]$\tabularnewline
\hline 
\hline 
\multirow{3}{*}{$14$} & $30$ & $55.35$\tabularnewline
\cline{2-3} 
 & $100$ & $13.28$\tabularnewline
\cline{2-3} 
 & $200$ & $1.73$\tabularnewline
\hline 
\multirow{3}{*}{$15.5$} & $100$ & $57.11$\tabularnewline
\cline{2-3} 
 & $200$ & $32.43$\tabularnewline
\cline{2-3} 
 & $500$ & $5.94$\tabularnewline
\hline 
\multirow{3}{*}{$17$} & $1000$ & $63.86$\tabularnewline
\cline{2-3} 
 & $3000$ & $26.02$\tabularnewline
\cline{2-3} 
 & $8000$ & $2.76$\tabularnewline
\hline 
\multirow{3}{*}{$18.5$} & $1\negthickspace\times\negthickspace10^{7}$ & $55.05$\tabularnewline
\cline{2-3} 
 & $2\negthickspace\times\negthickspace10^{7}$ & $30.30$\tabularnewline
\cline{2-3} 
 & $7\negthickspace\times\negthickspace10^{7}$ & $1.53$\tabularnewline
\hline 
\end{tabular}%
\begin{tabular}{|c|c|c|}
\hline 
\multicolumn{3}{|c|}{$\gamma=0.8$}\tabularnewline
\hline 
\hline 
$h$ & $r$ & $\frac{C(r)}{C(1)}[\%]$\tabularnewline
\hline 
\hline 
\multirow{3}{*}{$11$} & $20$ & $52.90$\tabularnewline
\cline{2-3} 
 & $50$ & $19.36$\tabularnewline
\cline{2-3} 
 & $100$ & $3.62$\tabularnewline
\hline 
\multirow{3}{*}{$12.4$} & $100$ & $47.52$\tabularnewline
\cline{2-3} 
 & $200$ & $22.42$\tabularnewline
\cline{2-3} 
 & $400$ & $4.99$\tabularnewline
\hline 
\multirow{3}{*}{$12.8$} & $200$ & $65.86$\tabularnewline
\cline{2-3} 
 & $500$ & $35.09$\tabularnewline
\cline{2-3} 
 & $1000$ & $12.29$\tabularnewline
\hline 
\multirow{3}{*}{$13$} & $2000$ & $57.37$\tabularnewline
\cline{2-3} 
 & $4000$ & $32.91$\tabularnewline
\cline{2-3} 
 & $1\negthickspace\times\negthickspace10^{4}$ & $6.21$\tabularnewline
\hline 
\end{tabular}
\end{table}

In Table~\ref{tab:P2} the ratio of correlation functions given by
$\frac{C_{I}(r)}{C_{I}(1)}=\frac{C_{IH}(r)}{C_{IH}(1)}=\frac{C_{H}(r)}{C_{H}(1)}=\frac{C(r)}{C(1)}=\left(\frac{\lambda_{-}}{\lambda_{+}}\right)^{r-1}$
in percent is reported as a function of the assumed distance $r$
for various fixed magnetic fields and their corresponding pseudo-critical
temperature $T_{p}$. Here we can see that as the magnetic field increases,
the system shows strong correlation between distant spins. For $\gamma=0.75$
and $h=18.5$ the correlation function weakened compared to its nearest
neighbor in about 50\% for $r\approx10^{7}$. Certainly, this is completely
unexpected compared to the standard one-dimensional spin chain.

Additional plots of magnetizations and correlation functions for $\gamma=0.7$
and $\gamma=0.8$ are reported in Appendix~\ref{sec:app-b}, where
we observe similar behavior as for $\gamma=0.75$.

\begin{figure}[h]
\centering{}\includegraphics[scale=0.3]{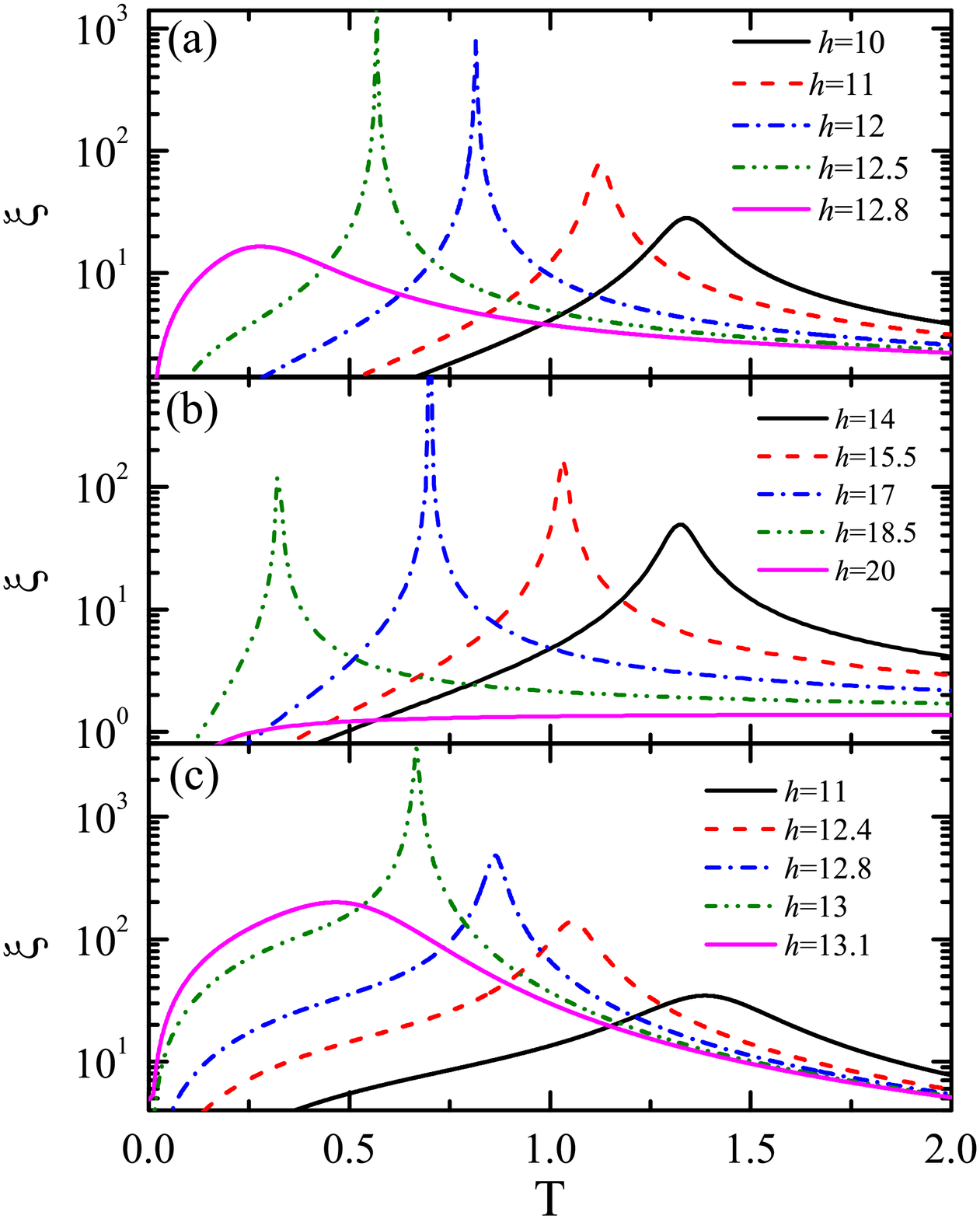} \caption{\label{fig:corrlt-lengh}Correlation length against temperature, for
the parameters considered in Fig.~\ref{fig:2}. (a) $\gamma=0.7$;
(b) $\gamma=0.75$; (c) $\gamma=0.8$.}
\end{figure}

In addition, here we will discuss the correlation length $\xi(T)=\left(\ln\frac{\lambda_{+}}{\lambda_{-}}\right)^{-1}$,
which characterizes exponential decay of correlations with distance.
In Fig.~\ref{fig:corrlt-lengh}(a) the correlation length $\xi(T)$
is depicted assuming $\gamma=0.7$ and the same parameters as considered
in Fig.~\ref{fig:2}. We observe the peaks that are located at the
pseudo-critical temperature $T_{p}$. The correlation length becomes
extremely large, but remains finite at the pseudo-critical temperature
$T_{p}$, the lower the temperature is, the larger the correlation
length is. For higher temperatures the peak becomes smaller and wider.
Analogously, Fig.~\ref{fig:corrlt-lengh}(b) and Fig.~\ref{fig:corrlt-lengh}(c)
refer to $\gamma=0.75$ and $\gamma=0.8$; we observed here similar
behavior.

In principle, the internal energy can be obtained using the relation
\begin{equation}
\mathcal{U}=\langle H\rangle=-\frac{\partial\lambda_{+}}{\partial\beta}.\label{eq:U_in}
\end{equation}
Alternatively, the one-cell correlations are related to thermodynamics,
since they determine the internal energy $\mathcal{U}$ of the spin
system by 
\begin{alignat}{1}
\mathcal{U}=\langle H\rangle= & -J(1+\gamma)\langle S_{a}^{x}S_{b}^{x}\rangle-J(1-\gamma)\langle S_{a}^{y}S_{b}^{y}\rangle\nonumber \\
 & -J_{z}\langle S_{a}^{z}S_{b}^{z}\rangle-4J_{0}\langle S^{z}\sigma\rangle-2h_{z}\langle S^{z}\rangle-h\langle\sigma\rangle.
\end{alignat}
Using the previous results \eqref{eq:SS_alph}, \eqref{eq:avr-Szsg-alt},
\eqref{eq:<sz>} and \eqref{eq:<sg>}, we find an equivalent result
to that obtained from \eqref{eq:U_in}.

\section{Conclusions}

\label{sec5}

To summarize, we have examined the properties of the spin-$\frac{1}{2}$
Ising-XYZ diamond chain in the regime where the model shows pseudo-transitions
and quasi-phases \cite{Souza2017}. These pseudo-transitions are not
true finite-temperature transitions but only sudden changes such as
in the entropy, internal energy, and magnetization, which are quite
similar to a first-order phase transition. While in some other thermodynamic
quantities, in such as the specific heat, magnetic susceptibility,
correlation length and correlation functions, sharp peaks arise which
is also quite similar to a second-order phase transition. Therefore,
this effect could be confused when interpreting experimental data
and misinterpreted as a true phase transition.

A simple way to understand the presence of quasi-phases and pseudo-critical
temperature could be the mapping of the original spin-$\frac{1}{2}$
Ising-XYZ diamond chain onto a simple effective ferromagnetic ($J_{\mathrm{eff}}>0$)
Ising model with effective magnetic field $h_{\mathrm{eff}}$ through
a decoration transformation. The zero effective magnetic field $h_{{\rm {eff}}}(T_{p})=0$
when $\bar{w}_{0}\rightarrow0$ indicates the presence of the so-called
pseudo-critical temperature $T_{p}$ leading to a simultaneous flip
of all Ising spins. Previously in Ref.~\cite{Souza2017} an equivalent
condition $w_{1}=w_{-1}$ when $\bar{w}_{0}\rightarrow0$ was considered.

Hence, we analyzed the quasi-phase diagram at low temperatures and
determined the region of parameters where the pseudo-transitions may
occur. Here we consider a detailed investigation of Ising spin and
Heisenberg spin magnetization, as well as the pair correlation function
with arbitrary distance. Basically in this model we have three types
of correlation functions: Ising-Ising spin correlation function $C_{I}$,
Ising-Heisenberg spin correlation functions $C_{IH}$, and Heisenberg-Heisenberg
spin correlation functions $C_{H}$. The magnetizations of the Ising
and Heisenberg spin illustrate the presence of a substantial change
in magnetization near the pseudo-critical temperature. Likewise, all
correlation functions were also focused around the pseudo-critical
temperature, where we observed prominent peaks at pseudo-critical
temperature and this effect is supported by the analytical results.
It is also worth mentioning that the correlation function at pseudo-critical
temperature $T_{p}$ has large correlation length. For example, for
$\gamma=0.8$ and the parameters considered in Fig.~\ref{fig:2}
with magnetic field bit below the critical magnetic field ($h<h_{c}$),
i.e., $h=18.5$, the correlation functions become almost insensitive
to $r$ for up to $r\sim10^{6}$.

\appendix

\section{Heisenberg spin matrices}

\label{sec:app-a}

In this appendix we give in detail the elements of some important
matrices. First, we consider for the Heisenberg spin at site $a$
(see Fig.~\ref{fig:0}): 
\begin{equation}
\hat{\mathbf{s}}_{a}^{z}=\frac{1}{2}\left(\begin{array}{cccc}
-\cos(2\theta_{\mu}) & 0 & 0 & -\sin(2\theta_{\mu})\\
0 & 0 & -1 & 0\\
0 & -1 & 0 & 0\\
-\sin(2\theta_{\mu}) & 0 & 0 & \cos(2\theta_{\mu})
\end{array}\right),
\end{equation}
where $\tan(2\theta_{\mu})=\frac{J\gamma}{2J_{0}\mu+2h}$ and $0<\theta_{\mu}<\pi$.
Similarly, the other components become 
\begin{equation}
\hat{\mathbf{s}}_{a}^{x}=\frac{1}{2}\left(\begin{array}{cccc}
0 & \cos\hat{\theta}_{\mu} & -\sin\hat{\theta}_{\mu} & 0\\
\cos\hat{\theta}_{\mu} & 0 & 0 & \sin\hat{\theta}_{\mu}\\
-\sin\hat{\theta}_{\mu} & 0 & 0 & \cos\hat{\theta}_{\mu}\\
0 & \sin\hat{\theta}_{\mu} & \cos\hat{\theta}_{\mu} & 0
\end{array}\right),
\end{equation}
by $\hat{\theta}_{\mu}$ we define conveniently $\hat{\theta}_{\mu}=\theta_{\mu}+\frac{\pi}{4}$,
and 
\begin{equation}
\hat{\mathbf{s}}_{a}^{y}=\frac{i}{2}\left(\begin{array}{cccc}
0 & \sin\hat{\theta}_{\mu} & -\cos\hat{\theta}_{\mu} & 0\\
-\sin\hat{\theta}_{\mu} & 0 & 0 & \cos\hat{\theta}_{\mu}\\
\cos\hat{\theta}_{\mu} & 0 & 0 & \sin\hat{\theta}_{\mu}\\
0 & -\cos\hat{\theta}_{\mu} & -\sin\hat{\theta}_{\mu} & 0
\end{array}\right).
\end{equation}

Second, we obtain similarly for the Heisenberg spin at site $b$ (see
Fig.~\ref{fig:0}): 
\begin{equation}
\hat{\mathbf{s}}_{b}^{z}=\frac{1}{2}\left(\begin{array}{cccc}
-\cos(2\theta_{\mu}) & 0 & 0 & -\sin(2\theta_{\mu})\\
0 & 0 & 1 & 0\\
0 & 1 & 0 & 0\\
-\sin(2\theta_{\mu}) & 0 & 0 & \cos(2\theta_{\mu})
\end{array}\right),
\end{equation}
\begin{equation}
\hat{\mathbf{s}}_{b}^{x}=\frac{1}{2}\left(\begin{array}{cccc}
0 & \cos\hat{\theta}_{\mu} & \sin\hat{\theta}_{\mu} & 0\\
\cos\hat{\theta}_{\mu} & 0 & 0 & \sin\hat{\theta}_{\mu}\\
\sin\hat{\theta}_{\mu} & 0 & 0 & -\cos\hat{\theta}_{\mu}\\
0 & \sin\hat{\theta}_{\mu} & -\cos\hat{\theta}_{\mu} & 0
\end{array}\right),
\end{equation}
and 
\begin{equation}
\hat{\mathbf{s}}_{b}^{y}=\frac{i}{2}\left(\begin{array}{cccc}
0 & \sin\hat{\theta}_{\mu} & \cos\hat{\theta}_{\mu} & 0\\
-\sin\hat{\theta}_{\mu} & 0 & 0 & \cos\hat{\theta}_{\mu}\\
-\cos\hat{\theta}_{\mu} & 0 & 0 & -\sin\hat{\theta}_{\mu}\\
0 & -\cos\hat{\theta}_{\mu} & \sin\hat{\theta}_{\mu} & 0
\end{array}\right).
\end{equation}

Therefore, $\tilde{w}_{1}^{z}$ is given more explicitly 
\begin{alignat}{1}
\tilde{w}_{1}^{z}= & \frac{h_{z}+J_{0}}{\Delta_{1}}{\rm e}^{\beta\left(\frac{h}{2}+\frac{J_{z}}{4}\right)}\sinh(\beta\Delta_{1})\cos^{2}\phi+\frac{h_{z}}{\Delta_{0}}{\rm e}^{\frac{\beta J_{z}}{4}}\sinh(\beta\Delta_{0})\sin(2\phi)\nonumber \\
 & +\frac{h_{z}-J_{0}}{\Delta_{-1}}{\rm e}^{\beta\left(-\frac{h}{2}+\frac{J_{z}}{4}\right)}\sinh(\beta\Delta_{-1})\sin^{2}\phi.
\end{alignat}
Using the the relation \eqref{eq:Id-trig} and assuming $\bar{w}_{0}\rightarrow0$,
we find the following expansions up to order $\bar{w}_{0}^{2}$, 
\begin{alignat}{1}
\cos^{2}\phi= & \frac{1}{2}\left(1+\frac{w_{1}-w_{-1}}{B}\right)=\frac{1}{2}\left[1+\frac{w_{1}-w_{-1}}{|w_{1}-w_{-1}|}\left(1-2\bar{w}_{0}^{2}+\mathcal{O}\left(\bar{w}_{0}^{4}\right)\right)\right],\label{A8}\\
\sin^{2}\phi= & \frac{1}{2}\left(1-\frac{w_{1}-w_{-1}}{B}\right)=\frac{1}{2}\left[1-\frac{w_{1}-w_{-1}}{|w_{1}-w_{-1}|}\left(1-2\bar{w}_{0}^{2}+\mathcal{O}\left(\bar{w}_{0}^{4}\right)\right)\right].\label{A9}
\end{alignat}

\section{Additional correlation quantities}

\label{sec:app-b}

\begin{figure}[t]
\centering{}\includegraphics[scale=0.43]{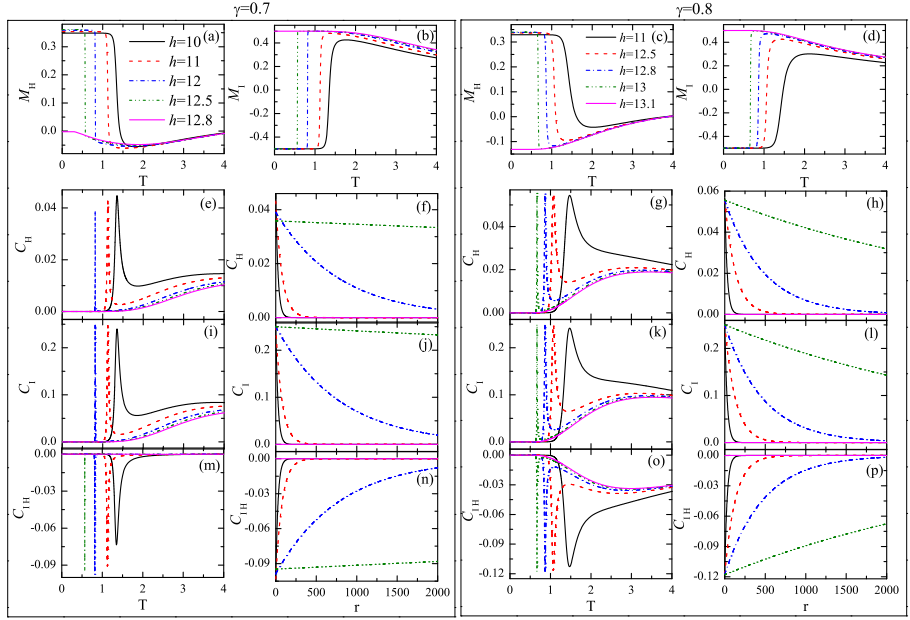} \caption{\label{fig:11}The left block of columns corresponds to fixed $\gamma=0.7$
and fixed $J=100$, $J_{z}=24$, and $J_{0}=-24$. (a) Heisenberg
spin magnetization as a function of temperature. (b) Ising spin magnetization
as a function of temperature. (e), (f) Heisenberg spin correlation
functions. (i), (j) Ising spin correlation functions. (m), (n) Ising-Heisenberg
spin correlation functions. Similarly, the right block of columns
corresponds to fixed $\gamma=0.8$. (c), (d) magnetizations as a function
of temperature. (g), (h) Heisenberg spin correlation functions. (k),
(l) Ising spin correlation functions. (o), (p) Ising-Heisenberg spin
correlation functions.}
\end{figure}

Here we report additional plots concerning magnetizations and correlation
functions, see Fig.~\ref{fig:11}. Mainly we observe similar behavior
as it was discussed in the main text. The only difference we worth
to mention would be that this pseudo-critical temperature occurs between
two quasi-phases: For $\gamma=0.7$ the pseudo-transition occurs between
$qFI$ and $qMF_{0}$, whereas for $\gamma=0.8$ the pseudo-transition
occurs between $qMF_{0}$ and $qMF_{2}$.

\section*{Acknowledgments}

I.~M.~Carvalho and J.~Torrico thank CAPES for full financial support.
S.~M.~de~Souza and O.~Rojas thank CNPq, CAPES and FAPEMIG for
partial financial support. O.~Derzhko was supported by the Brazilian
agency FAPEMIG (CEX - BPV-00090-17); he appreciates the kind hospitality
of the Federal University of Lavras in October-December of 2017.

\end{document}